\documentclass[11pt,letterpaper]{article}

\usepackage[utf8]{inputenc}
\usepackage[T1]{fontenc}
\usepackage[margin=1in]{geometry}
\usepackage{times}
\usepackage{cite}
\usepackage{amsmath,amssymb,amsfonts}
\usepackage{amsthm}
\usepackage{algorithm}
\usepackage{algorithmic}
\usepackage{graphicx}
\usepackage{booktabs}
\usepackage{multirow}
\usepackage{textcomp}
\usepackage{url}
\usepackage{bm}
\usepackage{hyperref}
\usepackage{subfig}
\usepackage{tikz}
\usepackage{authblk}
\usetikzlibrary{arrows.meta,positioning,fit,calc}

\graphicspath{{figuras/}{figures/}{./}}

\hypersetup{
  colorlinks=true,
  linkcolor=blue,
  citecolor=blue,
  urlcolor=blue
}

\def\BibTeX{{\rm B\kern-.05em{\sc i\kern-.025em b}\kern-.08em
    T\kern-.1667em\lower.7ex\hbox{E}\kern-.125emX}}

\begin{document}

	\title{Structure-Informed Estimation for Pilot-Limited MIMO Channels via Tensor Decomposition}
	
	\author[1]{Alexandre Barbosa de Lima}
	\affil[1]{Faculty of Exact Sciences and Technology, Pontifical Catholic University of São Paulo (PUC-SP), São Paulo, Brazil}
	\affil[ ]{\textit{Email: ablima@pucsp.br}}
	
	\date{\today}
	
	\maketitle
	
\begin{abstract}
Accurate channel state information in wideband multiple-input multiple-output (MIMO) systems is fundamentally constrained by pilot overhead, a challenge that intensifies as antenna counts and bandwidths scale toward 6G. This paper proposes a structure-informed hybrid estimator that formulates pilot-limited MIMO channel estimation as low-rank tensor completion from sparse pilot observations---a severely underdetermined inverse problem that prior tensor approaches avoid by assuming fully observed received signal tensors. Canonical polyadic~(CP) and Tucker decompositions are comparatively analyzed: CP excels for specular channels whose rank-one multipath structure matches the CP parameterization exactly, while Tucker provides greater numerical stability at extreme pilot scarcity where CP exhibits heavy-tail divergence. A lightweight three-dimensional U-shaped network (3D U-Net) learns residual components beyond the dominant low-rank structure, compensating for diffuse scattering and hardware non-idealities that algebraic priors alone cannot capture.
On synthetic specular channels, Tucker completion achieves $10.88$~dB normalized mean-squared error (NMSE) improvement over least squares and $7.83$~dB over orthogonal matching pursuit at $\rho = 10\%$ pilot density; CP outperforms Tucker by $13.11$~dB at signal-to-noise ratio (SNR)\,=\,20~dB under the specular multipath model. On DeepMIMO ray-tracing channels, the hybrid tensor--neural network (Tensor--NN) estimator splits into two complementary operating regimes: Tensor--NN(Tucker) remains stable at $\rho = 2\%$ where CP diverges, while a CP-guided ablation becomes the best-performing hybrid from $\rho \geq 4\%$, reaching $-16.44$~dB at $\rho = 8\%$ and $-20.34$~dB at $\rho = 20\%$. The Tucker-guided variant still outperforms unconstrained deep learning across the full pilot-density range, and the CP-guided variant widens that gap once the CP prior becomes stable. Empirical recovery threshold analysis confirms that sample complexity scales with intrinsic channel dimensionality---governed by the number of dominant propagation paths---rather than with the ambient tensor size.
\end{abstract}
	
	
\vspace{1em}
	
\section{Introduction}
\label{sec:intro}

Wideband multiple-input multiple-output (MIMO) systems with large antenna arrays are fundamental enablers of high-capacity wireless links in 5G and beyond networks. Point-to-point MIMO configurations are particularly relevant for wireless backhaul, fixed wireless access, vehicle-to-infrastructure communications, and satellite feeder links~\cite{marzetta2016fundamentals,heath2016overview}. However, the spectral efficiency gains from spatial multiplexing critically depend on accurate channel state information (CSI), whose acquisition becomes progressively challenging as antenna counts, bandwidth, and mobility increase~\cite{bjornson2017massive}.

Recent advances in radio-frequency (RF) integration have fundamentally altered the architectural landscape of large-array systems. Fully digital transceivers---once considered impractical at millimeter-wave frequencies---are now recognized as scalable and energy-efficient, with commercial deployments already underway at sub-6~GHz and expected at millimeter-wave (mmWave) frequencies~\cite{bjornson2019massive}. This architectural shift is further motivated by emerging 6G use cases such as Integrated Sensing and Communications (ISAC), where fine-grained, per-antenna access to received signals is essential for joint optimization across communication and sensing tasks~\cite{liu2022integrated}. The present article focuses on fully digital MIMO orthogonal frequency-division multiplexing (MIMO-OFDM) architectures with direct baseband access to individual antenna-domain channel coefficients.

Conventional estimation techniques such as least squares (LS) and linear minimum mean-square error (LMMSE) require pilot resources scaling with channel dimensionality. In wideband MIMO-OFDM systems, the channel comprises $N_r N_t N_f$ complex coefficients, where $N_r$, $N_t$, and $N_f$ denote the number of receive antennas, transmit antennas, and subcarriers, respectively, translating to tens of thousands of unknowns for systems with tens of antennas and hundreds of subcarriers~\cite{rao2014distributed,alkhateeb2014channel}. Deep learning addresses pilot scarcity by learning nonlinear mappings from sparse observations to full reconstructions~\cite{he2018deep,soltani2019deep,boloursaz2020deep,guo2024deep,gao2021deep}, achieving strong performance when training and deployment conditions match. However, these methods typically require large labeled datasets, offer limited interpretability, and generalize poorly under distribution shifts---limitations particularly pronounced in pilot-limited regimes. A fundamental challenge lies in the limited availability of representative training data: unlike computer vision or natural language processing (NLP), acquiring labeled physical-layer data from operational networks is costly and often impractical~\cite{bjornson2019massive}.

Physics-informed learning mitigates generalization issues by embedding prior knowledge through constraints or regularization~\cite{karniadakis2021physics}, often via ray tracing or site-specific geometric models~\cite{he2025attention,bock2025physics}. Although effective in controlled settings, these approaches depend on detailed environmental information and scale poorly across heterogeneous deployments. This paper adopts a complementary perspective: exploiting algebraic structure as a universal physical prior. Rather than encoding propagation physics through explicit geometry, algebraic constraints enforce low rank, separability, and multilinear structure arising from fundamental scattering mechanisms---properties that emerge broadly from sparse multipath propagation and array geometry, making them relevant as modeling priors across deployment scenarios~\cite{marzetta2016fundamentals,heath2016overview,alkhateeb2014channel}. In wideband systems, these properties combine into hierarchical spatial--frequency correlations naturally captured by tensor decompositions~\cite{sidiropoulos2017tensor,zhou2017lowrank,auddy2025tensors}.

A comparison between tensor completion and compressed sensing (CS) highlights key structural differences. CS-based approaches assume sparsity in a predefined angular--delay dictionary and recover coefficients from linear measurements, but are susceptible to basis mismatch in off-grid scenarios. Tensor completion instead exploits low-dimensional multilinear structure---such as canonical polyadic (CP) or Tucker rank---directly in the native antenna--frequency domain without explicit dictionary construction. These perspectives are complementary: sparsity captures parametric specular structure, while multilinear low-rank models capture broader correlations and offer robustness under richer scattering conditions.

Prior tensor-based methods have become increasingly relevant to wireless inference problems, as reviewed in~\cite{chen2021tensor}. In channel estimation, Zhou et al.~\cite{zhou2017lowrank} formulated mmWave estimation as CP decomposition of a fully observed received-signal tensor after RF combining, i.e., an overdetermined setting tailored to hybrid architectures rather than pilot-limited completion. Tensor formulations have also been explored for massive MIMO-OFDM via tensorized minimum mean-square error (MMSE) and orthogonal matching pursuit (OMP) estimators that exploit multilinear sparse structure~\cite{araujo2019tensor}, for intelligent reflecting surface (IRS)-assisted MIMO via parallel factor analysis (PARAFAC) modeling of cascaded channels under structured pilot and phase-shift patterns~\cite{de2021channel}, and for mmWave 3-D MIMO-OFDM via Estimation of Signal Parameters via Rotational Invariance Techniques (ESPRIT)-oriented tensor-train decomposition~\cite{gong2023esprit}. Tensor methods have further been applied to MIMO relay systems through nested Tucker decompositions~\cite{rocha2019closed} and to time-varying channels under high mobility~\cite{zhang2022tensor}. More recently, deep learning has also been integrated into tensor inference itself, for example by using neural networks to generate favorable initializations for canonical polyadic alternating least squares (CP-ALS) in massive MIMO channel estimation~\cite{gong2024deep}; beyond wireless-specific formulations, generic low-rank tensor completion has likewise been strengthened by deep priors through plug-and-play learning strategies~\cite{zhao2020deep}. These works collectively confirm the value of multilinear structure and learning-enhanced tensor inference, but they generally rely on received-signal tensor models, structured cascaded channels, or parametric decomposition settings rather than addressing pilot-limited recovery as low-rank completion directly on a partially observed channel tensor. This paper addresses that complementary scenario in fully digital MIMO-OFDM systems, where only a sparse subset of channel tensor entries is observed, and further studies the accuracy--stability trade-off between CP and Tucker priors under sparse pilot observations.

The main contributions of this article are:
\begin{enumerate}
	\item \textbf{Tensor completion formulation for pilot-limited
		MIMO:}
	Unlike prior methods assuming fully observed
	tensors~\cite{zhou2017lowrank}, this work formulates
	pilot-limited MIMO channel estimation as low-rank tensor
	completion from sparse pilot observations---a severely
	underdetermined inverse problem that requires structural
	priors for reliable reconstruction.
	
	\item \textbf{Hybrid tensor--neural architecture:}
	A two-stage tensor--neural network (Tensor--NN) estimator combining low-rank tensor completion
	with neural residual learning. The Tucker-guided variant
	remains stable at $\rho = 2\%$ where CP exhibits
	heavy-tail divergence (mean normalized mean-squared error (NMSE) $\approx +13.96$~dB),
	while the CP-guided ablation becomes the best-performing
	hybrid from $\rho \geq 4\%$, outperforming
	Tensor--NN(Tucker) by $2.72$~dB at $\rho = 8\%$ and
	$1.90$~dB at $\rho = 20\%$.
	
	\item \textbf{Comprehensive validation:}
	Evaluation against LS, LMMSE, OMP, and a purely data-driven residual network (ResNet)
	baseline demonstrates 10.88~dB NMSE improvement
	over LS and 7.83~dB over OMP at $\rho = 10\%$ on
	synthetic channels; on DeepMIMO ray-tracing channels,
	Tensor--NN(Tucker) consistently outperforms unconstrained
	deep learning across the full pilot-density range, while
	Tensor--NN(CP) yields further gains once the CP prior
	becomes stable.
\end{enumerate}

The remainder of this paper is organized as follows. Section~\ref{sec:system} introduces the system and channel models. Section~\ref{sec:tensor} presents tensor representations and algebraic priors. Section~\ref{sec:framework} describes the estimation framework. Section~\ref{sec:results} presents simulation results. Section~\ref{sec:conclusion} concludes the paper.

\section{System Model}
\label{sec:system}

This paper denotes scalars by lowercase letters (e.g., $\alpha$), vectors by bold lowercase letters (e.g., $\mathbf{a}$), matrices by bold uppercase letters (e.g., $\mathbf{A}$), and tensors by calligraphic letters (e.g., $\mathcal{H}$). The wideband MIMO channel takes the form of a third-order tensor $\mathcal{H} \in \mathbb{C}^{N_r \times N_t
	\times N_f}$, where $\mathbf{w}[k]$ denotes the $k$-th entry of a
vector $\mathbf{w}$. The symbol $\circ$ denotes the outer product, $\times_n$ the $n$-mode product, $\|\cdot\|_F$ the Frobenius norm, $\odot$ the Hadamard (element-wise) product, $\otimes$ the Kronecker product, and $(\cdot)^H$ the conjugate transpose (adjoint). All tensor dimensions and ranks follow this convention throughout the paper.

Let $\mathbf{H}[k] \in \mathbb{C}^{N_r \times N_t}$ denote the frequency-domain channel matrix at subcarrier $k \in \{1, \ldots,N_f\}$. The received signal satisfies
\begin{equation}
	\mathbf{y}[k] = \mathbf{H}[k]\mathbf{x}[k] + \mathbf{n}[k],
	\label{eq:received_signal}
\end{equation}
where $\mathbf{x}[k] \in \mathbb{C}^{N_t}$ is the transmitted signal vector and $\mathbf{n}[k] \sim \mathcal{CN}(\mathbf{0}, \sigma^2 \mathbf{I})$ denotes additive white Gaussian noise~\cite{heath2018foundations}.

\subsection{Pilot-Based Channel Estimation}

During the training phase, known pilot symbols occupy a subset of resource elements $(i,j,k)$ of the channel tensor, so observations cover only a subset of its coefficients. Stacking the frequency-domain channel matrices $\mathbf{H}[k] \in \mathbb{C}^{N_r \times N_t}$ across all $N_f$ subcarriers yields the wideband MIMO channel as a third-order tensor
\begin{equation}
	\mathcal{H} \in \mathbb{C}^{N_r \times N_t \times N_f}.
	\label{eq:channel_tensor}
\end{equation}

Pilot-based acquisition covers only the subset of tensor entries indexed by
\begin{equation}
	\Omega \subset \{1, \ldots, N_r\} \times \{1, \ldots, N_t\}
	\times \{1, \ldots, N_f\}.
\end{equation}
The resulting observation model reads~\cite{zhou2017lowrank}
\begin{equation}
	\mathcal{Y}_\Omega = \mathcal{P}_\Omega(\mathcal{H} + \mathcal{N})
	= \mathcal{P}_\Omega(\mathcal{H}) + \mathcal{P}_\Omega(\mathcal{N}),
	\label{eq:observation_model}
\end{equation}
where $\mathcal{P}_\Omega(\cdot)$ denotes the entry-wise sampling operator in \eqref{eq:sampling_operator}
\begin{equation}
	[\mathcal{P}_\Omega(\mathcal{H})]_{ijk} =
	\begin{cases}
		\mathcal{H}_{ijk}, & \text{if } (i,j,k) \in \Omega, \\
		0,                 & \text{otherwise},
	\end{cases}
	\label{eq:sampling_operator}
\end{equation}
and $\mathcal{N}$ is independent and identically distributed (i.i.d.) complex Gaussian noise. In practical pilot-aided OFDM training, the channel tensor admits no direct observation. At each pilot location $(i,j,k) \in \Omega$, a known pilot symbol provides a noisy scalar observation of the corresponding channel coefficient $\mathcal{H}_{ijk}$. Accordingly, $\mathcal{P}_\Omega(\mathcal{H})$ collects these observations as entry-wise samples of the channel tensor corrupted by additive noise. This abstraction reduces the estimation problem to entry-wise sampling of the channel tensor with noise, which is standard in tensor completion and isolates the impact of pilot sparsity from waveform-specific effects. This article defines the signal-to-noise ratio (SNR) with respect to the average power per channel coefficient over the full tensor $\mathcal{H}$, independent of the pilot set $\Omega$ and pilot ratio $\rho$, as formalized in~\eqref{eq:snr_def}. This ensures strict SNR comparability across all pilot densities. Equivalently, $\mathcal{P}_\Omega(\mathcal{H}) = \mathcal{M} \odot \mathcal{H}$, where $\mathcal{M} \in \{0,1\}^{N_r \times N_t \times N_f}$ is a binary mask with $[\mathcal{M}]_{ijk} = 1$ if $(i,j,k) \in \Omega$. Stacking the per-subcarrier noise vectors $\mathbf{n}[k]$ across receive antennas and frequency indices yields the noise tensor $\mathcal{N}$, consistently with~\eqref{eq:received_signal}.

The tensor completion formulation treats pilot-limited observation as entry-wise sampling over an index set $\Omega$. The experiments draw $\Omega$ from a pilot mask with either random sampling, a regular grid, or a comb pattern in frequency. Random uniform sampling serves as the default model, consistent with standard practice in tensor completion literature~\cite{liu2012tensor, candes2010matrix}. Structured masks define deterministic index sets that satisfy the same observation model but may exhibit different empirical recovery behavior; their performance is assessed experimentally in Experiment~4.

When the pilot ratio ($\rho$) satisfies
\begin{equation}
	\rho = \frac{|\Omega|}{N_r N_t N_f} \ll 1,
	\label{eq:pilot_ratio}
\end{equation}
only a small fraction of channel tensor coefficients admits direct observation. In this pilot-limited regime, recovering the full channel tensor $\mathcal{H}$ from partial and noisy observations $\mathcal{Y}_\Omega$ constitutes a severely ill-posed inverse problem: unknowns vastly exceed measurements, the observation operator is highly rank-deficient, and the solution is non-unique and noise-sensitive~\cite{rao2014distributed,alkhateeb2014channel,candes2012exact}. This fundamental limitation motivates incorporating structural priors that exploit intrinsic spatial and spectral correlations of the channel to regularize the estimation problem and enable reliable reconstruction.

\subsection{Multipath Channel Model}

In sparse scattering environments---such as outdoor deployments with limited reflectors or higher-frequency bands---wireless propagation is typically dominated by a small number of specular paths, leading to low-rank multipath channels~\cite{heath2016overview}. Under this assumption, the frequency-domain channel at subcarrier $k$ admits the geometric representation
\begin{equation}
	\mathbf{H}[k] = \sum_{\ell=1}^{L} \alpha_\ell \,
	\mathbf{a}_r(\theta_\ell) \, \mathbf{a}_t^H(\phi_\ell) \,
	e^{-j2\pi k \tau_\ell}, \quad k = 1, \ldots, N_f,
	\label{eq:multipath}
\end{equation}
where $L$ denotes the number of dominant propagation paths, $\alpha_\ell\in \mathbb{C}$ are complex path gains, $\mathbf{a}_r(\theta_\ell)$ and $\mathbf{a}_t(\phi_\ell)$ are receive and transmit array steering vectors associated with angles of arrival and departure, respectively, assuming a 2D propagation model with azimuth-only angular parameters, and $\tau_\ell \in [0,1)$ represents the normalized propagation delay. The exponential term models frequency-dependent phase rotation induced by path delay across OFDM subcarriers.

Eq.~\eqref{eq:multipath} reveals that each propagation path contributes a separable outer-product structure across receive, transmit, and frequency modes. Consequently, the wideband channel tensor takes the form of a sum of rank-one components, yielding an intrinsic low-rank CP representation governed by the number of dominant paths~\cite{zhou2017lowrank,kolda2009tensor}. This physically grounded low-rank structure directly motivates tensor-based channel modeling and underpins the structure-informed estimation framework developed in Section \ref{sec:tensor}.

\section{Tensor Representations and Algebraic Priors}
\label{sec:tensor}

This section formalizes how fundamental propagation mechanisms in wideband MIMO channels manifest as intrinsic algebraic structure. It introduces CP and Tucker decompositions as principled parameterizations of this structure, providing physically interpretable and computationally efficient priors for pilot-limited channel estimation.

\subsection{Canonical Polyadic Decomposition}
\label{subsec:CP-decomp}

The sparse multipath channel model in~\eqref{eq:multipath} admits a natural representation in terms of the CP decomposition~\cite{kolda2009tensor},
\begin{equation}
	\mathcal{H} = \sum_{r=1}^{R} \mathbf{u}_r \circ \mathbf{v}_r
	\circ \mathbf{w}_r,
	\label{eq:cp_decomposition}
\end{equation}
where $\mathbf{u}_r \in \mathbb{C}^{N_r}$, $\mathbf{v}_r \in
\mathbb{C}^{N_t}$, and $\mathbf{w}_r \in \mathbb{C}^{N_f}$ are factor vectors associated with the receive, transmit, and frequency modes, respectively, and $R$ denotes the CP rank.

For channels with $L$ dominant propagation paths, selecting $R = L$ yields a parsimonious and physically interpretable representation, with the correspondence
\begin{equation}
	\mathbf{u}_\ell = \alpha_\ell \mathbf{a}_r(\theta_\ell), \quad
	\mathbf{v}_\ell = \mathbf{a}_t^*(\phi_\ell), \quad
	[\mathbf{w}_\ell]_k = e^{-j2\pi k \tau_\ell},
	\label{eq:cp_physical}
\end{equation}
where $k$ indexes the subcarrier frequencies. Absorbing the complex gain $\alpha_\ell$ into the receive-mode factor $\mathbf{u}_\ell$ is a matter of convention; the gain could equivalently enter $\mathbf{v}_\ell$, $\mathbf{w}_\ell$, or distribute across multiple factors. This scaling ambiguity is intrinsic to the CP decomposition: under the Kruskal uniqueness condition, the CP model is unique only up to permutation and scaling of its rank-one components~\cite{kolda2009tensor}.

This correspondence makes explicit that each rank-one tensor component represents a single physical propagation path, implying that the effective CP rank is governed by the number of dominant paths rather than the ambient channel dimensionality. As a result, CP is particularly well matched to propagation regimes dominated by specular ray-like components, such as millimeter-wave channels around 28~GHz, where a small number of strong paths typically captures most of the channel energy. This property underlies the effectiveness of CP-based models for sparse-channel and large-scale MIMO estimation~\cite{zhou2017lowrank}. The Kruskal uniqueness condition guarantees factor identifiability up to permutation and scaling, which suffices for reliable channel reconstruction. Extracting explicit physical parameters, such as angles of arrival or departure, from the estimated factors would require additional structural constraints---e.g., array calibration or explicit enforcement of Vandermonde structure---beyond the scope of this work, which focuses on CSI reconstruction rather than direct parameter estimation~\cite{sidiropoulos2017tensor}.

\subsection{Tucker Decomposition}
\label{subsec:tucker}

While CP enforces strict rank-one separability across all modes, the Tucker decomposition~\cite{delathauwer2000multilinear} provides a more flexible multilinear representation,
\begin{equation}
	\mathcal{H} = \mathcal{G} \times_1 \mathbf{U}_r \times_2
	\mathbf{U}_t \times_3 \mathbf{U}_f,
	\label{eq:tucker_decomposition}
\end{equation}
where $\mathcal{G} \in \mathbb{C}^{R_r \times R_t \times R_f}$ is a low-dimensional core tensor capturing inter-mode coupling. The integers $(R_r, R_t, R_f)$ denote the multilinear ranks of $\mathcal{H}$ along the receive, transmit, and frequency modes, respectively. The factor matrices $\mathbf{U}_r \in \mathbb{C}^{N_r \times R_r}$,
$\mathbf{U}_t \in \mathbb{C}^{N_t \times R_t}$, and $\mathbf{U}_f \in
\mathbb{C}^{N_f \times R_f}$ contain orthonormal columns spanning the dominant subspaces associated with each mode.

Tucker generalizes CP, which the formulation recovers when the core tensor $\mathcal{G}$ is superdiagonal with $R_r = R_t = R_f$, by relaxing strict separability across modes. The core tensor captures inter-mode interactions that independent rank-one components cannot express, enabling representation of richer spatial--spectral correlations. From a physical perspective, this flexibility allows Tucker models to capture clustered multipath and moderately diffuse scattering, where multiple rays share correlated angular and delay characteristics. The multilinear ranks $(R_r, R_t, R_f)$ thus control a trade-off between compression and fidelity: smaller ranks yield stronger dimensionality reduction, while larger ranks enable finer modeling of scattering richness and model mismatch.

\begin{figure}[!t]
	\centering
	\includegraphics[width=0.55\columnwidth]{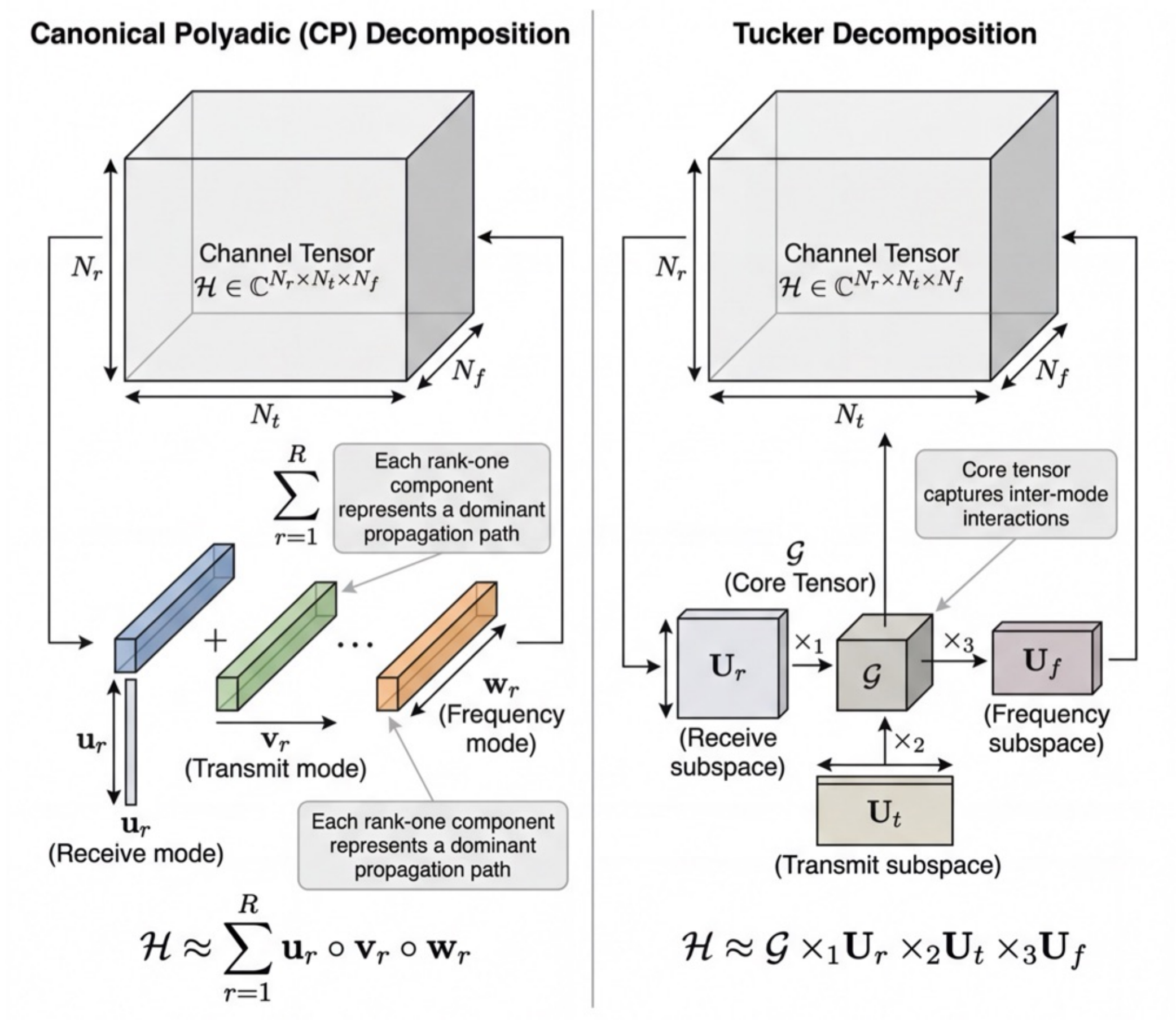}
	\caption{CP versus Tucker decompositions of the channel tensor
		$\mathcal{H} \in \mathbb{C}^{N_r \times N_t \times N_f}$.
		\textbf{Left:} CP expresses $\mathcal{H}$ as a sum of $R$
		rank-one components ($\mathbf{u}_r \circ \mathbf{v}_r \circ
		\mathbf{w}_r$), enforcing strict mode separability.
		\textbf{Right:} Tucker factorizes $\mathcal{H}$ into a core
		tensor $\mathcal{G}$ and mode matrices
		($\mathcal{G} \times_1 \mathbf{U}_r \times_2 \mathbf{U}_t
		\times_3 \mathbf{U}_f$), capturing inter-mode correlations.}
	\label{fig:cp_tucker}
\end{figure}

Fig.~\ref{fig:cp_tucker} illustrates the structural difference between CP and Tucker decompositions.

\subsection{Algebraic Priors from Propagation Physics}
\label{subsec:algebra-priors}

The tensor decompositions introduced above encode algebraic properties that directly reflect the physical mechanisms governing channel formation:
\begin{itemize}
	\item \textbf{Low rank}, reflecting the presence of a limited
	number of dominant propagation paths;
	\item \textbf{Structured factor matrices}, capturing array-induced
	spatial correlations;
	\item \textbf{Frequency smoothness}, arising from finite delay
	spread and limited coherence bandwidth.
\end{itemize}

The literature has extensively exploited the low-rank property through matrix completion and matrix sensing formulations for mmWave and terahertz (THz) systems~\cite{masood2023low}, complementing the tensor-based approach this paper adopts. Unlike geometric priors that rely on site-specific environmental descriptions or ray-tracing information, these algebraic constraints arise intrinsically from propagation physics and array structure, and therefore remain meaningful modeling assumptions across diverse deployment scenarios. As a result, tensor representations provide robust, interpretable, and physically grounded priors for pilot-limited channel estimation, consistent with prior tensor-based formulations in wireless communications~\cite{zhou2017lowrank}.

\section{Structure-Informed Learning Framework}
\label{sec:framework}

\begin{figure}[!t]
	\centering
	\includegraphics[width=0.98\textwidth]{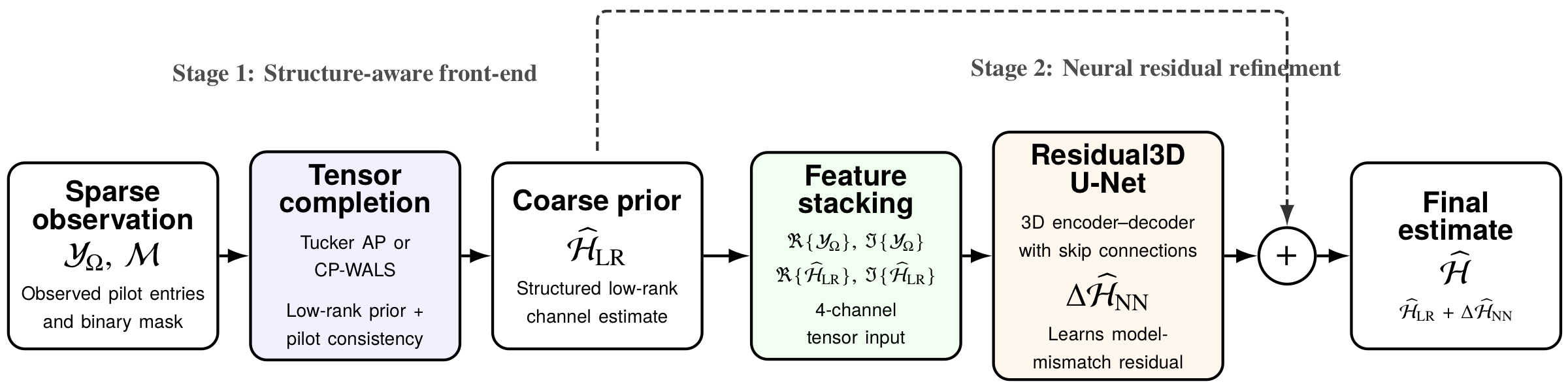}
	\caption{Detailed schematic of the Tensor--NN framework. Stage~1 obtains a coarse low-rank estimate $\widehat{\mathcal{H}}_{\mathrm{LR}}$ from the sparse observation tensor $\mathcal{Y}_{\Omega}$ and sampling mask $\mathcal{M}$ using either Tucker alternating projection or canonical polyadic weighted alternating least squares (CP-WALS). Stage~2 stacks the real and imaginary channels of $\mathcal{Y}_{\Omega}$ and $\widehat{\mathcal{H}}_{\mathrm{LR}}$, processes them with Residual3D U-Net, and adds the predicted residual $\Delta\widehat{\mathcal{H}}_{\mathrm{NN}}$ to produce the final channel estimate $\widehat{\mathcal{H}}$. Here, $\Re\{\cdot\}$ and $\Im\{\cdot\}$ denote the real and imaginary parts, respectively.}
	\label{fig:tensor_nn}
\end{figure}

This section presents the proposed structure-informed learning framework for pilot-limited MIMO channel estimation. The framework combines algebraic priors induced by propagation physics with data-driven residual learning in a unified estimation pipeline. A low-rank tensor completion stage first enforces intrinsic multilinear structure arising from sparse multipath propagation, array geometry, and frequency coherence, yielding a physically consistent coarse channel estimate. A lightweight neural network then learns residual components that the idealized low-rank model does not capture, such as diffuse scattering effects, hardware non-idealities, and model mismatch. By decoupling structure enforcement from residual learning, the proposed approach improves robustness in realistic pilot-limited regimes while retaining interpretability and computational efficiency. Fig.~\ref{fig:tensor_nn} illustrates the proposed two-stage estimation pipeline.

\subsection{Problem Formulation}
Given sparse pilot observations
\begin{equation}
	\mathcal{Y}_\Omega = \mathcal{P}_\Omega(\mathcal{H} + \mathcal{N})
	= \mathcal{P}_\Omega(\mathcal{H}) + \mathcal{P}_\Omega(\mathcal{N}),
	\label{eq:observation_model_framework}
\end{equation}
the goal is to estimate the unknown channel tensor $\mathcal{H}$ from incomplete and noisy measurements. This work poses the problem as
\begin{equation}
	\begin{split}
		\widehat{\mathcal{H}} = \;&\arg\min_{\mathcal{H}}
		\left\| \mathcal{P}_\Omega(\mathcal{H})
		- \mathcal{Y}_\Omega \right\|_F^2 \\
		&\text{s.t.} \quad
		\mathrm{rank}_{\mathrm{ml}}(\mathcal{H})
		= (R_r, R_t, R_f),
	\end{split}
	\label{eq:regularized_problem}
\end{equation}
where $\mathcal{P}_\Omega(\cdot)$ denotes projection onto the observed index set $\Omega$, $\mathcal{N}$ is i.i.d.\ complex Gaussian noise, and $\mathrm{rank}_{\mathrm{ml}}(\mathcal{H})$ denotes the multilinear rank of $\mathcal{H}$.

\subsection{Tensor Completion via Alternating Projection}

The alternating projection scheme enforces low-rank structure by iterating between: (i) projection onto a low-rank Tucker manifold via truncated higher-order singular value decomposition (HOSVD)~\cite{delathauwer2000multilinear}, and (ii) projection onto the data-consistency set defined by the observed pilot entries. This procedure alternates between structural regularization and measurement agreement.

Because the same low-rank front-end feeds both the tensor-only baselines and the hybrid estimators, Algorithm~\ref{alg:framework_pipeline} first summarizes the complete inference pipeline used throughout the paper: Tucker or CP coarse estimation, feature construction for the neural stage, residual refinement with the deployed Residual3DUNet, and the standalone ResNet branch used as an unconstrained baseline in Experiment~3. Algorithm~\ref{alg:tucker_completion} then details the Tucker completion subroutine invoked when the selected low-rank prior is Tucker. Let $\mathcal{Y}$ denote the observed tensor (zero-filled at unobserved entries) and
$\mathcal{M} \in \{0,1\}^{N_r \times N_t \times N_f}$ the sampling mask, where $[\mathcal{M}]_{ijk} = 1$ if $(i,j,k) \in \Omega$. The objective is to estimate $\widehat{\mathcal{H}}$ with multilinear rank $(R_r, R_t, R_f)$ consistent with the observations. Iterations terminate when $\delta^{(t)} < \epsilon$ or $t = T$; unless otherwise stated, $\epsilon = 10^{-6}$ and $T = 20$. For the tensor-guided hybrid front-end, convergence was empirically observed within $10$--$15$ iterations, so $T = 20$ provides sufficient margin without introducing unnecessary computation. The standalone Tucker baseline in Experiment~3 is run more conservatively with $T = 30$ to keep a uniform completion budget across the full pilot-ratio sweep, although convergence usually occurs earlier. The tolerance $\epsilon = 10^{-6}$ ensures that residual changes on the observed entries are negligible relative to the pilot-set norm, and was verified to be neither too tight nor too loose across the SNR and pilot-density ranges of all experiments.

Alternating projection defines a non-convex feasibility procedure that typically converges to a stationary point under standard regularity conditions~\cite{liu2012tensor},
without guarantees of global optimality; CP-based formulations exhibit analogous behavior. The framework assumes the multilinear ranks $(R_r, R_t, R_f)$ (Tucker)
and the CP rank $R$ known or selected a priori. In practice, rank selection may rely on information criteria (e.g., Akaike Information Criterion (AIC)) or singular-value inspection of tensor unfoldings~\cite{kolda2009tensor}. Synthetic experiments match ranks to the number of propagation paths ($L = 5$), whereas DeepMIMO ranks receive conservative choices based on representative singular-value decay. Underestimated ranks increase the residual $\Delta\mathcal{H} = \mathcal{H} - \mathcal{H}_{\text{LR}}$, motivating the neural residual stage (subsection~\ref{exp3}).

The Tucker stage performs reconstruction via alternating projections, iterating truncated HOSVD on the current estimate followed by enforcement of consistency on the observed entries (Algorithm~\ref{alg:tucker_completion}). The CP stage performs reconstruction via missing-data-aware CP-WALS, minimizing a weighted Frobenius norm over the observed entries only. The implementation optionally supports multiple random restarts, selecting the best solution by observed-data fit (NMSE on the pilot set) to reduce initialization sensitivity; when only one restart is used, this selection step is degenerate. Tucker alternating projections maintain a dense iterate and enforce data consistency on the observed set at each iteration, rather than optimizing exclusively on the sparse observation set. In the hybrid Tensor--NN pipeline, the low-rank reconstruction runs first and its output serves as the input prior for neural residual correction; for compactness, Algorithm~\ref{alg:framework_pipeline} keeps the CP branch at the level of its standard CP-WALS solver call, while Algorithm~\ref{alg:tucker_completion} expands the default Tucker branch step by step.

\begin{algorithm}[!t]
	\caption{Complete Structure-Informed Estimation Pipeline}
	\label{alg:framework_pipeline}
	\begin{algorithmic}
		\renewcommand{\algorithmicrequire}{\textbf{Input:}}
		\renewcommand{\algorithmicensure}{\textbf{Output:}}
		\REQUIRE Observed tensor $\mathcal{Y}$, mask $\mathcal{M}$,
		low-rank prior $p \in \{\text{Tucker}, \text{CP}, \text{None}\}$
		\REQUIRE Neural backend $b \in \{\text{None},
		\text{Residual3DUNet}, \text{ResNet}\}$
		\REQUIRE Tucker ranks $(R_r, R_t, R_f)$, CP rank $R$,
		iteration budgets $(T_{\mathrm{Tk}}, T_{\mathrm{CP}})$,
		tolerance $\epsilon$, network weights
		$(\boldsymbol{\phi}, \boldsymbol{\psi})$
		\ENSURE Final estimate $\widehat{\mathcal{H}}$
		\IF{$p = \text{Tucker}$}
		\STATE $\mathcal{H}_{\mathrm{LR}} \leftarrow
		\textsc{TuckerAP}(\mathcal{Y}, \mathcal{M},
		R_r, R_t, R_f, T_{\mathrm{Tk}}, \epsilon)$
		\ELSIF{$p = \text{CP}$}
		\STATE $(\mathbf{A}, \mathbf{B}, \mathbf{C}) \leftarrow
		\textsc{CP-WALS}(\mathcal{Y}, \mathcal{M},
		R, T_{\mathrm{CP}})$
		\STATE $\mathcal{H}_{\mathrm{LR}} \leftarrow
		\textsc{CP-Reconstruct}(\mathbf{A}, \mathbf{B}, \mathbf{C})$
		\ELSE
		\STATE $\mathcal{H}_{\mathrm{LR}} \leftarrow \mathbf{0}$
		\ENDIF
		\IF{$b = \text{None}$}
		\STATE $\widehat{\mathcal{H}} \leftarrow
		\mathcal{H}_{\mathrm{LR}}$
		\ELSIF{$b = \text{Residual3DUNet}$}
		\STATE $\mathcal{Z} \leftarrow
		\textsc{Concat}(\Re\{\mathcal{Y}\}, \Im\{\mathcal{Y}\},
		\Re\{\mathcal{H}_{\mathrm{LR}}\},
		\Im\{\mathcal{H}_{\mathrm{LR}}\})$
		\STATE $\Delta\mathcal{H}_{\mathrm{NN}} \leftarrow
		\textsc{Residual3DUNet}(\mathcal{Z}; \boldsymbol{\phi})$
		\STATE $\widehat{\mathcal{H}} \leftarrow
		\mathcal{H}_{\mathrm{LR}} + \Delta\mathcal{H}_{\mathrm{NN}}$
		\ELSE
		\STATE $\mathcal{Z} \leftarrow
		\textsc{Concat}(\Re\{\mathcal{Y}\}, \Im\{\mathcal{Y}\},
		\mathcal{M})$
		\STATE $\widehat{\mathcal{H}} \leftarrow
		\textsc{ResNet3D}(\mathcal{Z}; \boldsymbol{\psi})$
		\ENDIF
		\RETURN $\widehat{\mathcal{H}}$
	\end{algorithmic}
\end{algorithm}

\begin{algorithm}[!t]
	\caption{Tucker Completion via Alternating Projection}
	\label{alg:tucker_completion}
	\begin{algorithmic}
		\renewcommand{\algorithmicrequire}{\textbf{Input:}}
		\renewcommand{\algorithmicensure}{\textbf{Output:}}
		\REQUIRE Observed tensor $\mathcal{Y}$, mask $\mathcal{M}$,
		ranks $(R_r, R_t, R_f)$, max iterations $T$, tolerance
		$\epsilon$
		\ENSURE Completed tensor $\widehat{\mathcal{H}}$
		\STATE $\mathcal{X}^{(0)} \leftarrow \mathcal{Y}$
		\FOR{$t = 1, \ldots, T$}
		\STATE $\widehat{\mathcal{X}} \leftarrow
		\text{HOSVD}_{\text{trunc}}(\mathcal{X}^{(t-1)},
		R_r, R_t, R_f)$
		\STATE $\mathcal{X} \leftarrow \mathcal{M} \odot \mathcal{Y}
		+ (1 - \mathcal{M}) \odot \widehat{\mathcal{X}}$
		\STATE $\delta^{(t)} \leftarrow
		\frac{\|(\mathcal{X}-\mathcal{X}^{(t-1)})
			\odot\mathcal{M}\|_F}
		{\|\mathcal{Y}\odot\mathcal{M}\|_F+\varepsilon}$
		\IF{$\delta^{(t)} < \epsilon$}
		\STATE \textbf{break}
		\ENDIF
		\STATE $\mathcal{X}^{(t)} \leftarrow \widehat{\mathcal{X}}$
		\ENDFOR
		\RETURN $\widehat{\mathcal{H}} \leftarrow \widehat{\mathcal{X}}$
	\end{algorithmic}
\end{algorithm}

\subsection{Computational Complexity}

For CP decomposition with rank $R$, the number of structural degrees of freedom scales as $\mathcal{O}(R(N_r + N_t + N_f))$, compared with $\mathcal{O}(N_r N_t N_f)$ for unstructured estimation~\cite{kolda2009tensor}. Tucker models with multilinear rank $(R_r,R_t,R_f)$ require $\mathcal{O}(N_r R_r + N_t R_t + N_f R_f + R_r R_t R_f)$ parameters. Both are far smaller than the ambient tensor size in pilot-limited settings. For the DeepMIMO setup of Experiment~3 ($N_r \times N_t \times N_f = 16 \times 32 \times 64$), the standalone Tucker baseline with ranks $(4,4,8)$ uses 832 structural degrees of freedom, while CP with rank $R=5$ uses 560, versus 32,768 coefficients for an unconstrained channel tensor.

Algorithmic scaling alone, however, does not describe practical cost. Table~\ref{tab:runtime_complexity} therefore reports measured wall-clock completion/inference times together with approximate giga floating-point operations (GFLOPs) and trainable parameters. The tensor methods incur no offline learning cost but require iterative completion at inference time. The learned models invert this trade-off: they require offline training, yet their online complexity is fixed after deployment. Both Tensor--NN variants share the same 60,164-parameter residual network and remain markedly smaller than the purely data-driven ResNet baseline (225,794 parameters), while preserving the structural prior responsible for their accuracy advantage. Peak memory is not tabulated separately because the mixed central processing unit (CPU)/Metal Performance Shaders (MPS) implementation on Apple unified memory does not expose a stable backend-resident breakdown that is directly comparable across iterative tensor solvers and neural inference; the most reproducible cross-method indicators are therefore parameters, approximate GFLOPs, and measured wall-clock times.

\begin{table*}[!t]
	\centering
	\caption{Measured computational cost on the Experiment~3 DeepMIMO setup ($16 \times 32 \times 64$, $\rho=8\%$, SNR $=10$~dB) using a MacBook Pro with Apple M3 Pro. Offline training times for learned models are estimated from measured full-epoch wall-clock durations under the actual $1500/300$ training protocol.}
	\label{tab:runtime_complexity}
	\begin{tabular}{@{}lrrrr@{}}
		\toprule
		\textbf{Method} & \textbf{Trainable params.} & \textbf{Approx. GFLOPs} & \textbf{Offline training} & \textbf{Online time/sample} \\
		\midrule
		Tucker only & 0 & 0.016 & --- & 409.2~ms \\
		CP only & 0 & 0.025 & --- & 828.0~ms \\
		Tensor--NN(Tucker) & 60,164 & 1.065 & 81.0~min & 286.8~ms \\
		Tensor--NN(CP) & 60,164 & 1.072 & 3.4~min (fine-tune) & 557.1~ms \\
		ResNet & 225,794 & 7.389 & 45.6~min & 17.0~ms \\
		\bottomrule
	\end{tabular}
	\vspace{1mm}
	\\{\footnotesize The reported Tensor--NN(Tucker) timing corresponds to the deployed tensor-guided checkpoint, whose low-rank front-end uses Tucker ranks $(4,4,4)$ with 20 alternating-projection iterations; the standalone Tucker baseline uses $(4,4,8)$ with 30 iterations, as in Experiment~3. Tensor--NN(CP) uses the same residual backbone with a CP-WALS front-end of rank $R{=}5$ and 30 weighted alternating least-squares iterations, whereas the standalone CP baseline uses $R{=}5$ and 50 iterations. The CP-guided variant is obtained by 3-epoch fine-tuning on a dedicated 100/20 adaptation split after replacing the Tucker front-end. The longer offline training time of Tensor--NN(Tucker) is dominated by per-sample Tucker completion inside the training loop, whereas ResNet trains directly from the observed tensors.}
\end{table*}

\subsection{Hybrid Tensor--Neural Extension}

Although tensor decompositions capture dominant low-dimensional structure induced by sparse multipath propagation, residual errors arise from diffuse scattering, hardware impairments, or model mismatch. To account for these effects, the channel takes the form
\begin{equation}
	\mathcal{H} = \mathcal{T}(\boldsymbol{\theta}) +
	\Delta\mathcal{H}_{\text{NN}}(\boldsymbol{\phi}),
	\label{eq:hybrid_model}
\end{equation}
where $\mathcal{T}(\boldsymbol{\theta})$ denotes a low-rank tensor component and $\Delta\mathcal{H}_{\text{NN}}(\boldsymbol{\phi})$ a lightweight neural residual. This decomposition restricts learning to a correction term rather than the full channel, reducing sample complexity and mitigating overfitting.

Estimation proceeds in two stages. First, the framework obtains the structured component $\mathcal{T}(\boldsymbol{\theta})$ by solving the rank-constrained problem \eqref{eq:regularized_problem} with a chosen low-rank solver---truncated HOSVD with alternating projections for Tucker, or weighted alternating least squares for CP. Second, the neural network learns the residual by minimizing the supervised mean-squared error (MSE) between predicted and true residuals,
\begin{equation}
	\mathcal{L}_{\text{NN}}(\boldsymbol{\phi}) =
	\left\| \Delta\mathcal{H}_{\text{NN}}(\boldsymbol{\phi}) -
	(\mathcal{H} - \mathcal{H}_{\text{LR}}) \right\|_F^2,
	\label{eq:nn_loss}
\end{equation}
where $\mathcal{H}_{\text{LR}}$ is the low-rank reconstruction and $\mathcal{H}$ the ground-truth channel. This two-stage approach aligns with model-driven and algorithm-unrolling paradigms in signal processing~\cite{gregor2010learning,monga2021algorithm,he2019model}. 
The two stages are trained sequentially: the tensor completion stage is executed first and its output $\mathcal{H}_{\text{LR}}$ is treated as fixed during neural network training,
with no gradient propagation through the tensor stage.
The low-rank stage is modular: Tucker completion uses alternating projections, whereas CP completion uses missing-data-aware weighted alternating least squares. The main hybrid reported in this work uses Tucker as the default structural prior because of its superior numerical stability at extreme pilot scarcity, while Section~\ref{exp3} also evaluates a CP-guided ablation to quantify the resulting accuracy--stability trade-off.

\subsubsection{Neural Architecture and Training}

\begin{figure}[t]
	\centering
	\includegraphics[width=0.75\columnwidth]{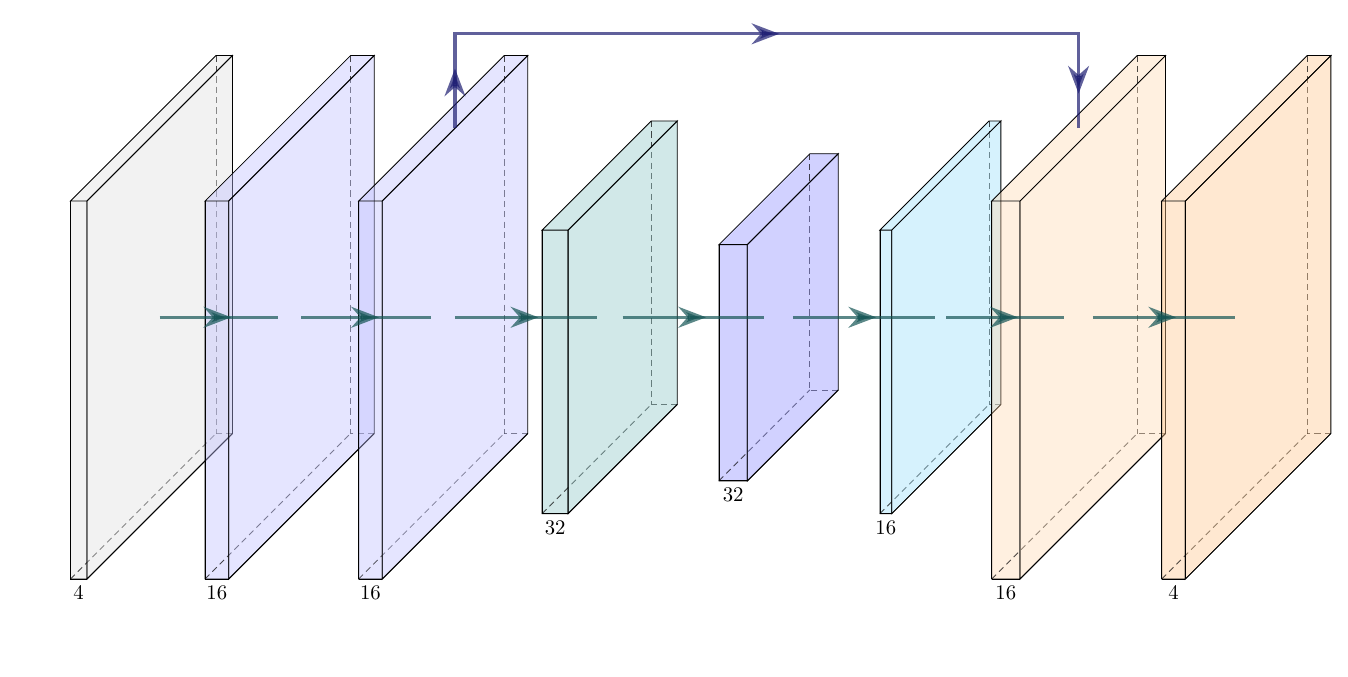}
	\caption{\textbf{Residual3DUNet architecture} ($C_{\text{in}} \in
		\{2,3,4,5\}$, base width $C = 16$). Encoder-1 applies two
		$\text{Conv3D}{+}\text{ReLU}\,(3^3/1)$ layers followed by
		$\text{Conv3D}\,(2^3/2)$ downsampling ($C{\to}2C$). Encoder-2
		keeps $2C{\to}2C$ at the bottleneck. The decoder upsamples via
		$\text{ConvTranspose3D}\,(2^3/2)$, concatenates encoder skip
		features, and maps $2C{\to}C$. A final $\text{Conv3D}\,(3^3/1)$
		projects $C{\to}C_{\text{in}}$, yielding $\Delta\mathcal{H}_{
			\mathrm{NN}}$.}
	\label{fig:residual3dunet}
\end{figure}

The residual term $\Delta\mathcal{H}_{\text{NN}}$ takes the form of a lightweight three-dimensional U-shaped network (3D U-Net) architecture~\cite{ronneberger2015}, denoted \textbf{Residual3DUNet}. The implementation uses three-dimensional convolution (Conv3D), transposed three-dimensional convolution (ConvTranspose3D), and rectified linear unit (ReLU) blocks. U-Net offers a strong inductive bias in reconstruction tasks, combining multiscale context aggregation with skip connections that preserve fine structural information. This design suits the learning of systematic residual components arising from model mismatch rather than generic denoising. The 3D formulation operates directly on the spatial--frequency tensor, enabling joint modeling of antenna and subcarrier correlations. A lightweight configuration limits parameter count and keeps the tensor model as the primary structural prior. Because the architecture is fully convolutional, the same layer pattern can be instantiated for different tensor sizes $N_r \times N_t \times N_f$ without architectural redesign; this flexibility does not by itself imply cross-scenario generalization, and retraining or fine-tuning may still be required under size or scenario shifts. Table~\ref{tab:unet_architecture} reports the deployed 4-channel DeepMIMO checkpoint with layerwise output shapes and parameter counts, while Fig.~\ref{fig:residual3dunet} sketches the topology.

\begin{table}[!t]
	\centering
	\caption{Residual3DUNet architecture for the deployed 4-channel DeepMIMO checkpoint used in Experiment~3.}
	\label{tab:unet_architecture}
	\begin{tabular}{@{}lllr@{}}
		\toprule
		\textbf{Stage} & \textbf{Operation} & \textbf{Output shape $(C,N_r,N_t,N_f)$} & \textbf{Params.} \\
		\midrule
		Input      & Observed tensor + low-rank prior                 & $(4,16,32,64)$ & 0 \\
		Encoder-1a & Conv3D + ReLU ($3^3$/1)                          & $(16,16,32,64)$ & 1,744 \\
		Encoder-1b & Conv3D + ReLU ($3^3$/1)                          & $(16,16,32,64)$ & 6,928 \\
		Downsample & Conv3D ($2^3$/2)                                 & $(32,8,16,32)$ & 4,128 \\
		Encoder-2  & ReLU $\to$ Conv3D $\to$ ReLU ($3^3$/1)          & $(32,8,16,32)$ & 27,680 \\
		Upsample   & ConvTranspose3D ($2^3$/2)                        & $(16,16,32,64)$ & 4,112 \\
		Decoder-1  & Concat $\to$ ReLU $\to$ Conv3D $\to$ ReLU ($3^3$/1) & $(16,16,32,64)$ & 13,840 \\
		Output     & Conv3D ($3^3$/1)                                 & $(4,16,32,64)$ & 1,732 \\
		\midrule
		Total      & ---                                               & --- & 60,164 \\
		\bottomrule
	\end{tabular}
	\vspace{1mm}
	\\{\footnotesize Notation: $k^3$/s denotes kernel size $k \times k
		\times k$ with stride $s$. The reported shapes and counts correspond to the deployed Experiment~3 checkpoint with $C_{\text{in}} = 4$ and base width $C = 16$; the alternative $C_{\text{in}} \in \{2,3,5\}$ variants of Fig.~\ref{fig:residual3dunet} keep the same topology and differ only in the first and last layer counts.}
\end{table}

The input consists of the observed tensor $\mathcal{Y}_\Omega$ represented by separate real and imaginary channels, optionally augmented with the binary pilot mask and/or the real and imaginary parts of the low-rank estimate $\mathcal{H}_{\mathrm{LR}}$, yielding $C_{\text{in}} \in \{2,3,4,5\}$. The principal DeepMIMO Tensor--NN checkpoint used in Experiment~3 is tensor-guided and therefore uses $C_{\text{in}} = 4$ (real/imaginary parts of $\mathcal{Y}_\Omega$ and $\mathcal{H}_{\mathrm{LR}}$). The architecture employs a single encoder--decoder level. The base channel width $C$ equals 16 in all Tensor--NN experiments unless otherwise stated.

The encoder applies two $3 \times 3 \times 3$ convolutional layers with padding 1, each followed by ReLU activation, and performs downsampling via a $2 \times 2 \times 2$ strided convolution doubling the channel dimension. The second encoder block begins with ReLU, followed by a $3 \times 3 \times 3$ convolution and another ReLU. The decoder mirrors this structure using transposed convolution for upsampling, concatenation with corresponding encoder features through a skip connection, a single $3 \times 3 \times 3$ convolution bracketed by ReLU activations, and a final projection layer back to the input channel dimension, producing the residual estimate $\widehat{\mathcal{R}}$. When necessary, trilinear interpolation aligns tensor dimensions prior to skip-connection concatenation. The architecture employs no batch normalization, dropout, or other explicit regularization layers.

The framework obtains the reconstructed channel as
\begin{equation}
	\widehat{\mathcal{H}} = \mathcal{H}_{\text{LR}} +
	\widehat{\mathcal{R}},
	\label{eq:reconstruction}
\end{equation}
where $\mathcal{H}_{\text{LR}}$ denotes the low-rank tensor approximation. The tensor-guided checkpoint used in Experiment~3 contains 60,164 trainable parameters.

\paragraph*{Training details}
Training uses the Adam optimizer~\cite{kingma2015adam} with fixed learning rate $10^{-3}$ and batch size 2. The procedure uses no learning-rate decay or warm-up. The implementation initializes network weights using PyTorch default schemes, corresponding to Kaiming/He uniform initialization for convolutional and transposed convolutional layers, with biases set to zero when present~\cite{paszke2019pytorch}.

For the DeepMIMO experiment, both Tensor--NN(Tucker) and the ResNet baseline are trained on 1,500 paired samples $(\mathcal{H},\mathcal{Y}_\Omega)$ and validated on 300 samples generated from a fixed random split of the 500 line-of-sight (LOS) users retained in the filtered \texttt{ASU\_Campus\_3p5} subset. The user-level split assigns 400 users to training and 100 to validation (split seed 123, base\_seed $=42$). Classical sample-level $k$-fold cross-validation is not used here because the training pipeline procedurally re-samples pilot masks and noise from the same underlying users; distributing such augmented samples across folds would mix realizations of the same users and would therefore not provide an independent generalization test. Model selection instead relies on the fixed user-level validation split and, for Tensor--NN(Tucker), early stopping. During training, the pilot ratio is sampled independently and non-stratified as $\rho \sim \mathcal{U}[0.05,0.20]$, while SNR is sampled as $\mathrm{SNR} \sim \mathcal{U}[0,20]$~dB; the reported curves in Fig.~\ref{fig:exp3_deepmimo} evaluate the same trained models at fixed test ratios $\rho \in \{2,4,6,8,10,15,20\}\%$ and SNR $=10$~dB. Notably, the test ratios $\rho = 2\%$ and $4\%$ lie below the training range, so performance at these points reflects extrapolation beyond the training distribution.

The principal Tensor--NN(Tucker) checkpoint uses the tensor-guided 4-channel input, pilot-consistency weight $\lambda_{\mathrm{pilot}}=0.1$, and early stopping with patience 5 and minimum improvement $10^{-4}$. Training stopped after 16 epochs, selecting epoch 11 as the deployed checkpoint. The ResNet baseline uses three input channels (real/imaginary parts of $\mathcal{Y}_\Omega$ plus the pilot mask), the same optimizer and multi-pilot/multi-SNR sampling schedule, and runs for 25 epochs without early stopping. Because the low-rank prior is recomputed for each training sample, the offline cost of Tensor--NN(Tucker) is dominated by per-sample Tucker completion rather than by the residual U-Net itself, which explains why ResNet trains faster despite its larger parameter count and nominal floating-point operation (FLOP) budget. The CP-guided ablation of Section~\ref{exp3} reuses the same Residual3DUNet backbone and fine-tunes it for 3 epochs on a dedicated 100-sample training / 20-sample validation adaptation split after replacing the Tucker front-end by CP-WALS with 30 weighted alternating least-squares iterations. Table~\ref{tab:nn_training_protocol} summarizes the optimization settings used for these learned estimators.

\begin{table*}[!t]
	\centering
	\caption{Training protocol for the learned estimators used in Experiment~3 (DeepMIMO).}
	\label{tab:nn_training_protocol}
	\small
	\begin{tabular}{@{}p{0.18\textwidth}p{0.24\textwidth}p{0.22\textwidth}p{0.24\textwidth}@{}}
		\toprule
		\textbf{Setting} & \textbf{Tensor--NN(Tucker)} & \textbf{ResNet} & \textbf{Tensor--NN(CP)} \\
		\midrule
		Input / role & 4 channels: $\Re\{\mathcal{Y}_\Omega\}$, $\Im\{\mathcal{Y}_\Omega\}$, $\Re\{\mathcal{H}_{\mathrm{LR}}\}$, $\Im\{\mathcal{H}_{\mathrm{LR}}\}$; residual learning & 3 channels: $\Re\{\mathcal{Y}_\Omega\}$, $\Im\{\mathcal{Y}_\Omega\}$, pilot mask; direct channel regression & Same 4-channel Residual3DUNet backbone as Tensor--NN(Tucker), but with a CP coarse prior \\
		Training / validation data & 1,500 / 300 paired samples from 400 / 100 LOS users & Same split and sample counts as Tensor--NN(Tucker) & 100 / 20 adaptation samples from the same DeepMIMO scenario \\
		Optimizer / learning rate / batch size & Adam, $10^{-3}$, batch size 2 & Adam, $10^{-3}$, batch size 2 & Adam, $10^{-3}$, batch size 2 \\
		Main loss & Residual MSE, $\|\Delta\mathcal{H}_{\mathrm{NN}} - (\mathcal{H} - \mathcal{H}_{\mathrm{LR}})\|_F^2$ & Channel MSE, $\|\widehat{\mathcal{H}} - \mathcal{H}\|_F^2$ & Same residual-MSE objective as Tensor--NN(Tucker) \\
		Pilot-consistency weight & $\lambda_{\mathrm{pilot}} = 0.1$ & $\lambda_{\mathrm{pilot}} = 0.1$ & $\lambda_{\mathrm{pilot}} = 0.1$ \\
		Epoch policy & Early stopping with patience 5 and min.\ improvement $10^{-4}$; 16 epochs executed, best epoch 11 retained & 25 fixed epochs, no early stopping & 3 fine-tuning epochs after swapping the Tucker front-end for CP \\
		Training sweep & $\rho \sim \mathcal{U}[0.05,0.20]$, $\mathrm{SNR} \sim \mathcal{U}[0,20]$~dB & Same as Tensor--NN(Tucker) & Same multi-pilot / multi-SNR schedule during adaptation \\
		Low-rank front-end during training & Tucker completion with ranks $(4,4,4)$ and 20 alternating-projection iterations & None & CP-WALS with rank $R{=}5$ and 30 weighted alternating least-squares iterations \\
		\bottomrule
	\end{tabular}
\end{table*}

\subsubsection{Residual Structure and Architectural Rationale}

After low-rank tensor reconstruction, the residual $\Delta\mathcal{H} = \mathcal{H} - \mathcal{H}_{\text{LR}}$ exhibits two dominant characteristics: (i) spatial and spectral smoothness induced by bandlimited propagation and finite delay spread, and (ii) localized components arising from diffuse scattering, hardware impairments, and rank mismatch. This structure naturally spans multiple spatial and frequency scales.

The U-Net architecture matches this residual structure well: its encoder captures coarse global context through progressive downsampling, while skip connections preserve fine-scale localized information. This design enables efficient reconstruction of multi-scale residual patterns with a small parameter budget, a property proven effective in inverse problems and denoising tasks involving structured residuals.

\section{Simulation Results}
\label{sec:results}

This section evaluates pilot-limited wideband MIMO channel estimation under synthetic and ray-tracing-based propagation models through four experiments. Experiment~1 benchmarks Tucker completion against LS, LMMSE, and OMP baselines across pilot densities under a controlled synthetic channel at fixed SNR. Experiment~2 isolates the performance gap between CP and Tucker decompositions as a function of SNR under the synthetic specular channel model. Experiment~3 evaluates the hybrid Tensor--NN framework on DeepMIMO ray-tracing data, assessing performance on a realistic non-synthetic channel dataset. Experiment~4 characterizes sample complexity by empirically mapping recovery thresholds in the $(L,\rho)$ plane. Across all scenarios, the goal is to quantify how intrinsic tensor structure---in particular, low multilinear rank induced by sparse scattering---enables reliable reconstruction of the wideband channel tensor $\mathcal{H}$ from noisy and incomplete pilot observations.

\subsection{Setup}
\label{subsec:setup}

For synthetic experiments, the setup represents the wideband channel as a third-order tensor $\mathcal{H} \in \mathbb{C}^{N_r \times N_t \times N_f}$ with $N_r = N_t = 32$ antennas and $N_f = 128$ subcarriers, generating it according to the multipath model in~\eqref{eq:multipath} with $L$ dominant paths. Angles satisfy $\theta_\ell, \phi_\ell \sim \mathcal{U}[-\pi/2, \pi/2]$ (angular spread $180^\circ$), complex gains $\alpha_\ell$ are drawn i.i.d.\ from $\mathcal{CN}(0,1)$ and normalized to unit total power, and normalized delays satisfy $\tau_\ell \sim \mathcal{U}[0,1)$. Uniform linear arrays (ULAs) with half-wavelength spacing serve at both ends~\cite{heath2016overview,alkhateeb2014channel}. The setup models pilot acquisition as entry-wise random sampling over an index set $\Omega$ with pilot ratio $\rho=|\Omega|/(N_r N_t N_f)$~\cite{candes2010matrix,liu2012tensor}. 
The model applies additive white Gaussian noise (AWGN) only to observed entries. The noise variance is set according to a global SNR reference,
\begin{equation}
	\sigma_n^2 = \frac{\|\mathcal{H}\|_F^2 / (N_r N_t N_f)}{10^{\mathrm{SNR}/10}},
	\label{eq:snr_def}
\end{equation}
where $\|\mathcal{H}\|_F^2 / (N_r N_t N_f)$ denotes the average power per channel coefficient over the full tensor, independent of the pilot set $\Omega$ and pilot ratio $\rho$.
This definition is applied identically across all experiments, including the DeepMIMO scenario (Experiment~3), ensuring strict SNR comparability across pilot densities.

For synthetic experiments, the channel model operates in a normalized discrete-frequency domain, where delays are specified in sample units, leaving bandwidth undefined. Since the evaluation targets reconstruction accuracy rather than physical delay interpretation, results remain invariant to bandwidth scaling.

The following estimators enter the comparison:
\begin{itemize}
	\item \textbf{LS}: baseline estimator setting unobserved entries to zero;
	\item \textbf{LMMSE}: linear MMSE estimator with covariance estimated from channel samples, used as a practical statistical reference baseline;
	\item \textbf{OMP}: joint sparse recovery using simultaneous OMP (SOMP) with shared support across subcarriers and a Kronecker ULA steering dictionary (receive/transmit). For notation consistency, this joint SOMP curve carries the label ``OMP'' in the plots~\cite{alkhateeb2014channel};
	\item \textbf{Tucker}: completion via alternating projection between truncated HOSVD with multilinear ranks $(R_r, R_t, R_f)$ and data-consistency enforcement on $\Omega$~\cite{delathauwer2000multilinear,liu2012tensor};
	\item \textbf{CP}: rank-$R$ CP completion with missing-data-aware weighted alternating least squares, selecting the best restart by observed-data fit~\cite{kolda2009tensor};
	\item \textbf{Tensor--NN(Tucker) / Tensor--NN(CP)}: proposed hybrid architectures combining low-rank tensor completion and a lightweight 3D U-Net for residual learning, differing only in the structural front-end.
\end{itemize}
The experiments measure performance by NMSE,
\begin{equation}
	\mathrm{NMSE} = \frac{\|\widehat{\mathcal{H}} -
		\mathcal{H}\|_F^2}{\|\mathcal{H}\|_F^2},
	\label{eq:nmse}
\end{equation}
averaging each result over Monte Carlo (MC) realizations. Table~\ref{tab:sim_params} summarizes experiment settings.

\begin{table*}[!t]
	\centering
	\caption{Simulation parameters.
		Exp.~4 employs 100 Monte Carlo realizations, as its objective
		is recovery-threshold (sample-complexity) characterization
		rather than statistical performance averaging.}
	\label{tab:sim_params}
	\begin{tabular}{@{}lcccc@{}}
		\toprule
		\textbf{Parameter} & \textbf{Exp.\ 1} & \textbf{Exp.\ 2} &
		\textbf{Exp.\ 3} & \textbf{Exp.\ 4} \\
		\midrule
		Sweep variable
		& Pilot ratio & SNR & Pilot ratio & Pilot ratio / $L$ paths \\
		Pilot ratio
		& 2--20\% & 8\% & 2--20\% & 1--25\% \\
		SNR (dB)
		& 10 & $-5$ to 20 & 10 & 10 \\
		$L$ paths
		& 5 & 5 & 5 & $\{2,3,5,8,10,15\}$ \\
		$N_r \times N_t \times N_f$
		& $32 \times 32 \times 128$ & $32 \times 32 \times 128$
		& $16 \times 32 \times 64$ & $32 \times 32 \times 128$ \\
		Standalone Tucker
		& $(4,4,6),\,T{=}20$ & $(5,5,6),\,T{=}20$ & $(4,4,8),\,T{=}30$ & $(L,L,L+1),\,T{=}20$ \\
		Standalone CP
		& --- & $R{=}5,\,T{=}50$ & $R{=}5,\,T{=}50$ & --- \\
		Hybrid front-end
		& --- & --- & Tucker $(4,4,4),\,T{=}20$; CP $R{=}5,\,T{=}30$ & --- \\
		Dataset
		& Synthetic & Synthetic & DeepMIMO & Synthetic \\
		DeepMIMO users
		& --- & --- & 500 LOS (400/100 train/val) & --- \\
		NN train/val samples
		& --- & --- & 1500 / 300 & --- \\
		NN training $\rho$ / SNR
		& --- & --- & 5--20\% / 0--20~dB & --- \\
		MC runs
		& 500 & 500 & 500 & 100 \\
		\bottomrule
	\end{tabular}
\end{table*}

Experiment~1 uses ranks $(4,4,6)$ to assess robustness under mild rank underestimation; Experiment~2 uses $(5,5,6)$ to match the channel rank more closely and isolate the CP--Tucker comparison at higher SNR.

This work also evaluates realistic propagation conditions using ray-tracing-based channels from the DeepMIMO dataset~\cite{alkhateeb2019deepmimo}. The experiments adopt the \texttt{ASU\_Campus\_3p5} scenario (DeepMIMO v4 pre-release), generated with Remcom Wireless InSite at 3.5~GHz.

Experiment~3 employs DeepMIMO channels generated with an $8\times4$ uniform planar array (UPA) at the base station (BS) and a $4\times4$ UPA at the user equipment (UE), reshaped into tensors with $N_r=16$ receive-antenna coefficients, $N_t=32$ transmit-antenna coefficients, and $N_f=64$ subcarriers over 50~MHz bandwidth, yielding $\mathcal{H}\in\mathbb{C}^{16\times 32\times 64}$ (Table~\ref{tab:sim_params}). The study trims the scenario to 500 valid LOS users after 10\% user subsampling and uses channels without path truncation, so low-rank structure emerges empirically rather than being imposed. During model development, training/validation splits are formed by random user subsampling with fixed seeds; the final 500-run evaluation samples channels from the same filtered LOS pool with independent seeds rather than a fixed held-out user partition. Strict spatial hold-out evaluation therefore remains future work. Channels are normalized to unit average power prior to
noise addition, so that $\|\mathcal{H}\|_F^2/(N_r N_t N_f) = 1$ and the noise variance reduces to $\sigma_n^2 = 10^{-\mathrm{SNR}/10}$, consistently with the global SNR definition in~\eqref{eq:snr_def}.

Experiment~4 adopts the same synthetic channel model with $N_r = N_t = 32$ and $N_f = 128$, varying the number of paths $L \in \{2, 3, 5, 8, 10, 15\}$ and sweeping pilot ratios from $1\%$ to $25\%$ across 14 sample points, with Tucker ranks set to $(L, L, L+1)$. Each result averages over 100 Monte Carlo realizations, and the analysis defines successful recovery as $\mathrm{NMSE} \leq 10^{-2}$.


\subsection{Experiment 1: Tensor Completion under Pilot Scarcity}
\label{exp1}

\begin{figure}[!t]
	\centering
	\includegraphics[width=0.95\columnwidth]{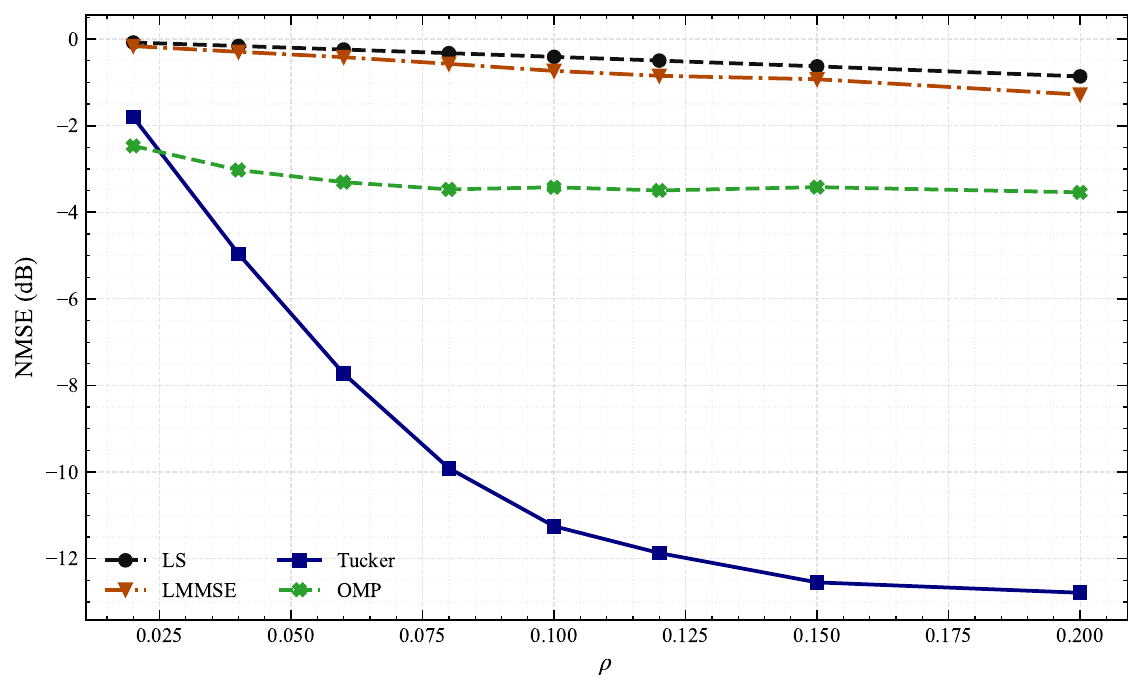}
	\caption{NMSE versus pilot ratio for synthetic specular channels (SNR $= 10$~dB, $L = 5$, Tucker ranks $(4,4,6)$, 500~MC runs). Tensor--NN is omitted because under the exact specular model its residual correction is visually indistinguishable from Tucker.}
	\label{fig:exp1_pilots}
\end{figure}

Fig.~\ref{fig:exp1_pilots} shows NMSE as a function of pilot ratio at SNR $= 10$~dB with $L = 5$ paths (500 MC runs, global-SNR setting). This experiment highlights the benefit
of Tucker tensor completion under pilot undersampling. CP is excluded from this comparison as its performance relative to Tucker under the synthetic specular channel
model is isolated in Experiment~2 (Section~\ref{exp2}), which sweeps SNR at fixed pilot ratio and provides a cleaner characterization of the CP--Tucker trade-off;
furthermore, CP exhibits heavy-tail divergence at low pilot densities (as confirmed in Experiment~3), which would compress the vertical scale and obscure the Tucker--baseline
comparison that is the focus of this experiment. The hybrid Tensor--NN estimator is also omitted here because, under the exact synthetic specular model, the residual term is negligible and the neural correction collapses onto the Tucker reconstruction, so plotting it would only duplicate the Tucker trajectory.

LS remains weak across the sweep, ranging from approximately $-0.08$~dB at $\rho = 2\%$ to $-0.86$~dB at $\rho = 20\%$, confirming that naive interpolation cannot recover channel structure from sparse pilots. LMMSE provides only modest improvement, decreasing from about $-0.16$~dB to $-1.28$~dB over the same range. OMP improves over LS/LMMSE and reaches around $-3.5$~dB at higher pilot ratios, but still exhibits a clear performance floor.

Tucker completion improves monotonically with pilot density, from about $-1.80$~dB at $\rho = 2\%$ to $-12.79$~dB at $\rho = 20\%$. At $\rho = 10\%$, Tucker achieves
$\mathrm{NMSE} \approx -11.25$~dB, corresponding to gains of $10.88$~dB over LS ($\approx -0.37$~dB), $10.52$~dB over LMMSE, and $7.83$~dB over OMP ($\approx -3.42$~dB). Notably, at the most extreme sparsity point ($\rho = 2\%$), OMP is slightly better than Tucker (about $0.67$~dB), but Tucker becomes clearly dominant from $\rho \geq 4\%$ onward.

\subsection{Experiment 2: CP versus Tucker Decomposition}
\label{exp2}

\begin{figure}[!t]
	\centering
	\includegraphics[width=0.95\columnwidth]{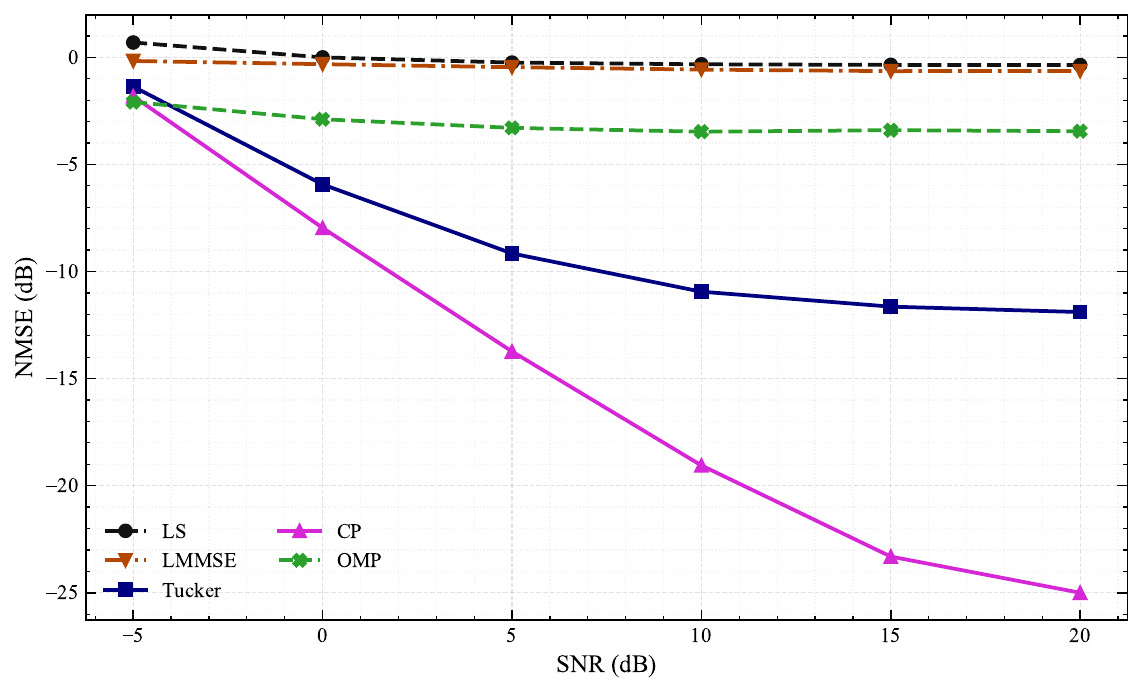}
	\caption{NMSE versus SNR for synthetic specular channels ($\rho = 8\%$, $L = 5$, Tucker ranks $(5,5,6)$, CP rank $R = 5$, 500~MC runs). Hybrid Tensor--NN curves are omitted because under this exact specular setting they collapse onto their underlying low-rank reconstructions.}
	\label{fig:exp2_cp_tucker}
\end{figure}

Fig.~\ref{fig:exp2_cp_tucker} compares Tucker and CP as a function of SNR at $\rho=8\%$. This experiment addresses decomposition choice under the synthetic specular channel model.

Both tensor methods outperform LS and LMMSE across the full SNR range, and CP consistently outperforms Tucker. At low SNR ($-5$~dB), CP already provides a slight gain over Tucker (CP $\approx -1.82$~dB versus Tucker $\approx -1.37$~dB, about $0.45$~dB). The gap widens markedly with SNR: at $20$~dB, CP reaches $\mathrm{NMSE}\approx -25.00$~dB versus Tucker $\approx -11.89$~dB, i.e., about $13.11$~dB advantage. Compared to OMP, Tucker is slightly worse at $-5$~dB (OMP $\approx -2.09$~dB) but becomes clearly better from $0$~dB onward. As in Experiment~1, the learned residual stage is omitted because the exact low-rank synthetic model leaves essentially no model-mismatch residual to correct.

This behavior is consistent with the synthetic model in~\eqref{eq:multipath}, where the channel consists of a small number of rank-one components. In this regime, CP provides a tighter representation than Tucker's full-core parameterization.

\subsection{Experiment 3: Hybrid Tensor--Neural Estimation on DeepMIMO}
\label{exp3}

\begin{figure}[!t]
	\centering
	\includegraphics[width=0.95\columnwidth]{%
		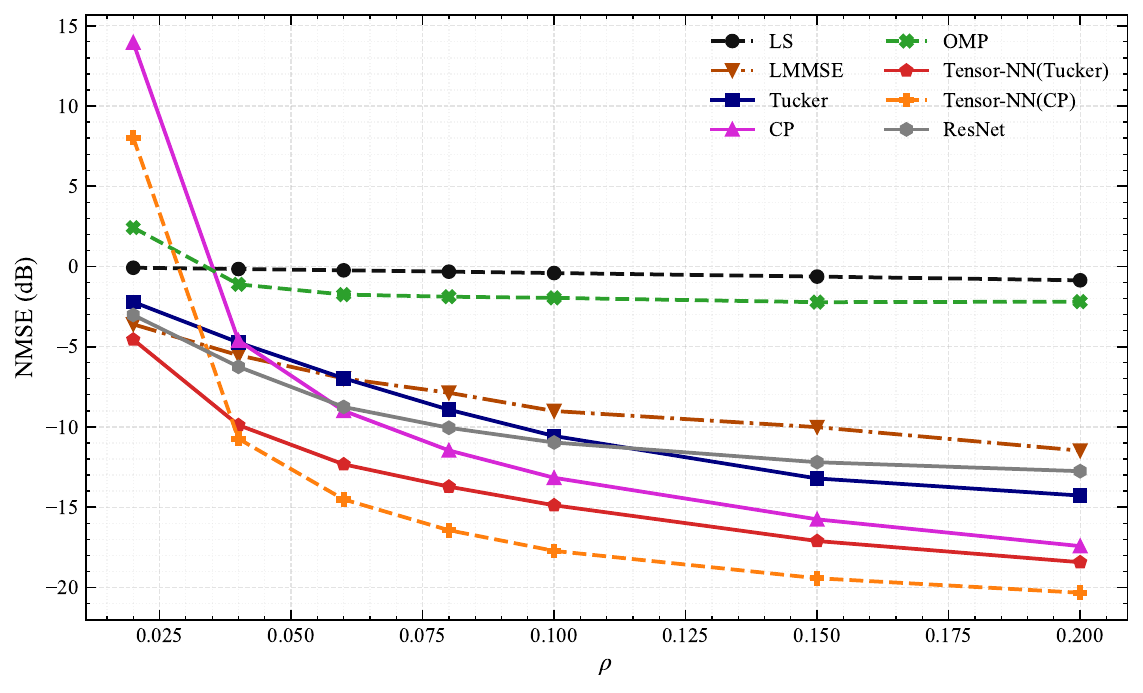}
	\caption{NMSE versus pilot ratio for DeepMIMO ASU\_Campus\_3p5 at SNR $= 10$~dB, comparing LS, LMMSE, OMP, Tucker, CP, Tensor--NN(Tucker), Tensor--NN(CP), and ResNet over 500 Monte Carlo runs.}
	\label{fig:exp3_deepmimo}
\end{figure}

Fig.~\ref{fig:exp3_deepmimo} evaluates the complete framework on ray-tracing-based DeepMIMO channels at SNR $= 10$~dB. These channels exhibit higher effective rank and stronger model mismatch relative to the idealized synthetic setting.

The comparison includes LS, LMMSE, OMP, Tucker, CP, Tensor--NN(Tucker), Tensor--NN(CP), and a purely data-driven ResNet baseline. The two hybrid variants split into complementary operating regimes. Tensor--NN(Tucker) is the most robust option at extreme pilot scarcity, with NMSE decreasing from approximately $-4.56$~dB at $\rho=2\%$ to approximately $-18.43$~dB at $\rho=20\%$. Because training samples only cover $\rho \in [5,20]\%$, the points $\rho=2\%$ and $4\%$ probe genuine extrapolation below the training range.

At the most severe undersampling point ($\rho=2\%$), raw CP diverges to approximately $+13.96$~dB and Tensor--NN(CP) remains strongly degraded at approximately $+8.03$~dB, whereas Tensor--NN(Tucker) stays well behaved. Once the CP prior leaves this divergent regime, however, Tensor--NN(CP) becomes the best-performing method from $\rho \geq 4\%$. It reaches approximately $-10.74$~dB, $-14.52$~dB, $-16.44$~dB, $-17.73$~dB, $-19.43$~dB, and $-20.34$~dB at $\rho=4\%$, $6\%$, $8\%$, $10\%$, $15\%$, and $20\%$, respectively.

At $\rho=8\%$, Tensor--NN(CP) improves by about $4.97$~dB over CP ($\approx -11.46$~dB), $2.72$~dB over Tensor--NN(Tucker) ($\approx -13.72$~dB), $7.51$~dB over Tucker ($\approx -8.93$~dB), and $8.57$~dB over LMMSE ($\approx -7.87$~dB). At $\rho=20\%$, Tensor--NN(CP) remains ahead by about $2.91$~dB over CP and $1.90$~dB over Tensor--NN(Tucker). The hybrid comparison is intentionally conservative with respect to the low-rank front-end: the standalone Tucker baseline uses ranks $(4,4,8)$ and 30 iterations, whereas Tensor--NN(Tucker) uses a more compressed Tucker prior $(4,4,4)$ and 20 iterations so that the residual U-Net must correct a larger structured error rather than benefiting from a stronger tensor stage. Likewise, Tensor--NN(CP) keeps the same CP rank $R{=}5$ but uses 30 weighted alternating least-squares iterations, while the standalone CP baseline is allotted 50 iterations to approach the best tensor-only solution; in the hybrid setting, the tensor stage only needs to provide a stable coarse prior for residual correction. These results clarify the rationale for using Tucker as the default structural prior of the framework: Tucker is safer when pilot density is extremely low, whereas CP becomes preferable once a modest pilot budget is available. Neural residual learning therefore mitigates, but does not fully eliminate, the stability--accuracy trade-off of the underlying low-rank model.

\paragraph*{Comparison with the ResNet baseline}

A purely data-driven ResNet baseline~\cite{he2016deep,burghal2023enhanced,barman2025csi} is included to quantify the contribution of tensor structural priors over unconstrained deep learning. The ResNet is a 3D architecture operating on the real and imaginary parts of the observation and the pilot mask (3-channel input), with a $3\!\times\!3\!\times\!3$ Conv3D stem followed by 4 residual blocks (each comprising two $3\!\times\!3\!\times\!3$ Conv3D layers with ReLU activations) and a final $3\!\times\!3\!\times\!3$ Conv3D predicting the 2-channel complex output (base channels $=32$). Training uses paired $(\mathcal{H},\mathcal{Y}_\Omega)$ samples optimized with Adam under multi-SNR and multi-pilot regimes, with pilot-consistency weight $\lambda_{\mathrm{pilot}}=0.1$.

Tensor--NN(Tucker) outperforms ResNet at all pilot ratios: by $1.53$~dB at $\rho=2\%$, $3.64$~dB at $\rho=4\%$, $3.58$~dB at $\rho=6\%$, $3.67$~dB at $\rho=8\%$, $3.91$~dB at $\rho=10\%$, $4.91$~dB at $\rho=15\%$, and $5.67$~dB at $\rho=20\%$. Tensor--NN(CP) widens this gap once CP becomes stable, reaching margins of $4.49$~dB at $\rho=4\%$, $6.38$~dB at $\rho=8\%$, and $7.58$~dB at $\rho=20\%$, although it is not robust enough at $\rho=2\%$ to replace Tucker as the default all-range configuration.

Overall, the hybrid Tensor--NN framework provides two complementary operating points under realistic DeepMIMO propagation conditions: Tucker-guided residual learning offers the safest full-range behavior, whereas CP-guided residual learning offers the best accuracy once the pilot budget is high enough to stabilize the CP prior.

Table~\ref{tab:exp3_summary} consolidates representative operating points from Fig.~\ref{fig:exp3_deepmimo}, making explicit how the preferred estimator changes between the extreme-scarcity regime ($\rho=2\%$) and the moderate-sparsity regime ($\rho\geq 8\%$).

\begin{table*}[!t]
	\centering
	\caption{Representative DeepMIMO NMSE values (dB) from Fig.~\ref{fig:exp3_deepmimo} at selected pilot ratios. More negative values are better.}
	\label{tab:exp3_summary}
	\begin{tabular}{@{}lccc@{}}
		\toprule
		\textbf{Method} & \textbf{$\rho = 2\%$} & \textbf{$\rho = 8\%$} & \textbf{$\rho = 20\%$} \\
		\midrule
		LS & $-0.08$ & $-0.32$ & $-0.86$ \\
		LMMSE & $-3.60$ & $-7.87$ & $-11.48$ \\
		OMP & $2.42$ & $-1.89$ & $-2.19$ \\
		Tucker & $-2.21$ & $-8.93$ & $-14.28$ \\
		CP & $13.96$ & $-11.46$ & $-17.43$ \\
		ResNet & $-3.03$ & $-10.05$ & $-12.76$ \\
		Tensor--NN(Tucker) & $\mathbf{-4.56}$ & $-13.72$ & $-18.43$ \\
		Tensor--NN(CP) & $8.03$ & $\mathbf{-16.44}$ & $\mathbf{-20.34}$ \\
		\bottomrule
	\end{tabular}
	\vspace{1mm}
	\\{\footnotesize Values correspond to the same 500-run evaluation plotted in Fig.~\ref{fig:exp3_deepmimo}.}
\end{table*}

\subsection{Experiment 4: Sample Complexity Analysis}
\label{exp4}

This experiment empirically characterizes the sample complexity of tensor completion by determining the recovery threshold of Tucker reconstruction as a function of the number of observed entries $|\Omega|$. The setup generates synthetic channels according to the model in Section~\ref{subsec:setup}, with $N_r = N_t = 32$ antennas, $N_f = 128$ subcarriers, and $L \in \{2, 3, 5, 8, 10, 15\}$ propagation paths. The sweep covers pilot ratios from 1\% to 25\% across 14 sample points, with finer spacing at low pilot ratios (and consequently finer resolution near the transition region). Each result averages over 100 Monte Carlo realizations, and the analysis defines successful recovery as $\mathrm{NMSE} \le 10^{-2}$, which determines the empirical threshold $|\Omega|_{\min}$. Following the implementation, the experiments adapt Tucker ranks to $L$ via $(R_r, R_t, R_f) = (L, L, L+1)$ (capped by $N_f$).

\begin{figure*}[!t]
	\centering
	\subfloat[SNR = 10 dB]{%
		\includegraphics[width=0.32\textwidth]{%
			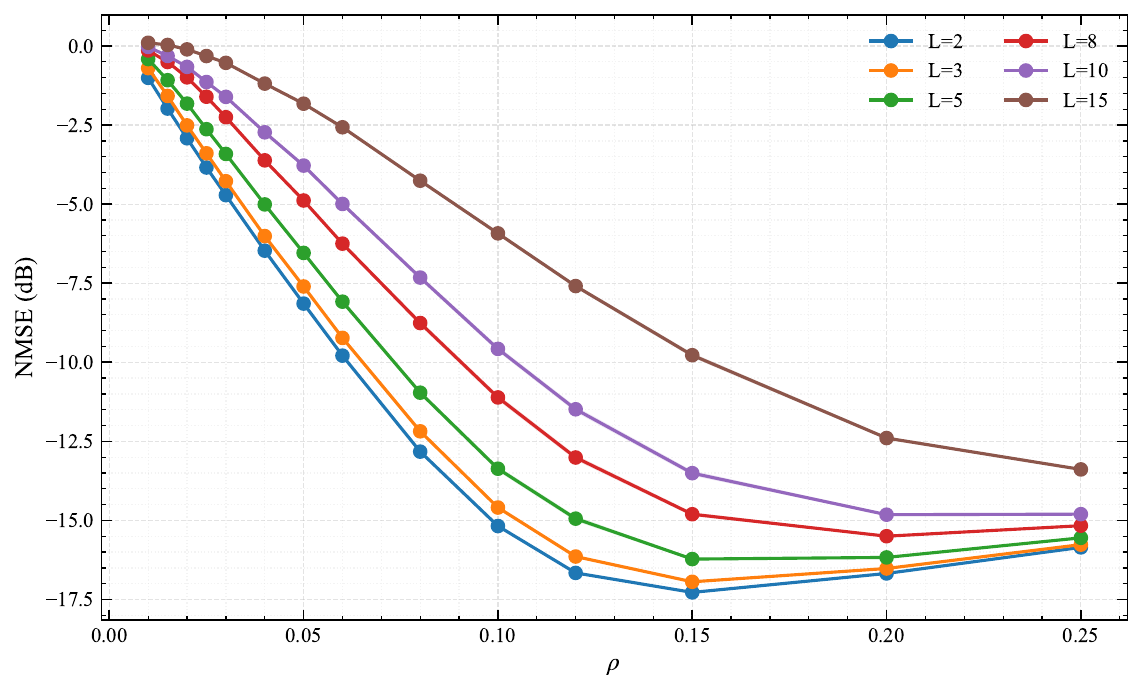}}
	\hfill
	\subfloat[SNR = 20 dB]{%
		\includegraphics[width=0.32\textwidth]{%
			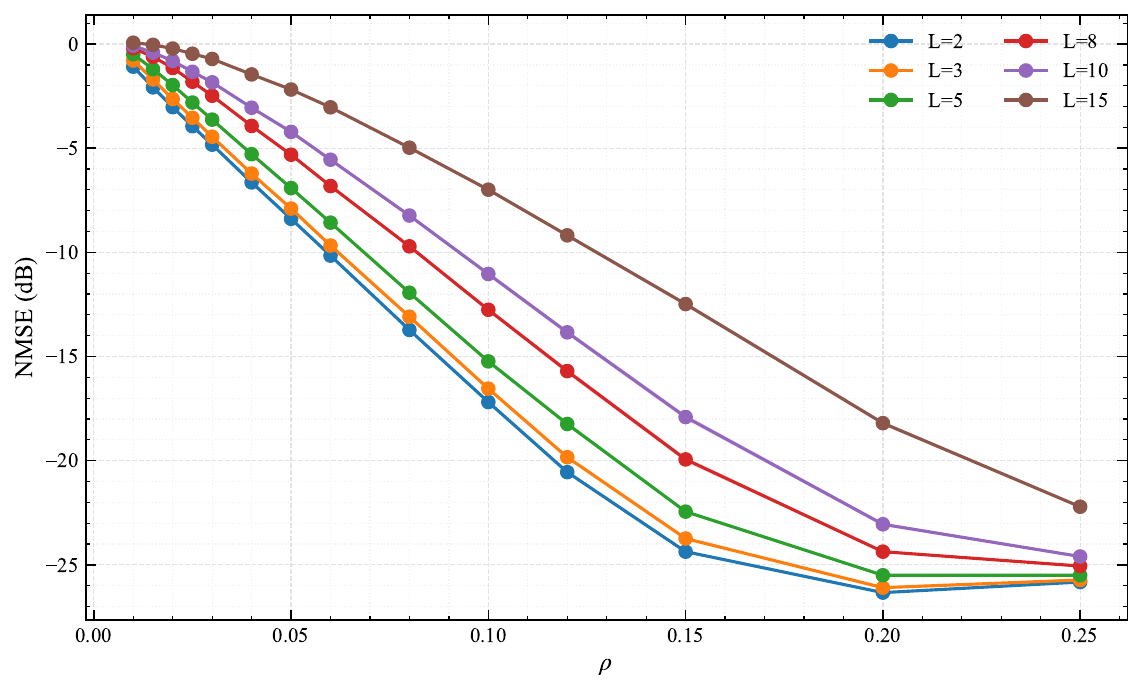}}
	\hfill
	\subfloat[SNR = 30 dB]{%
		\includegraphics[width=0.32\textwidth]{%
			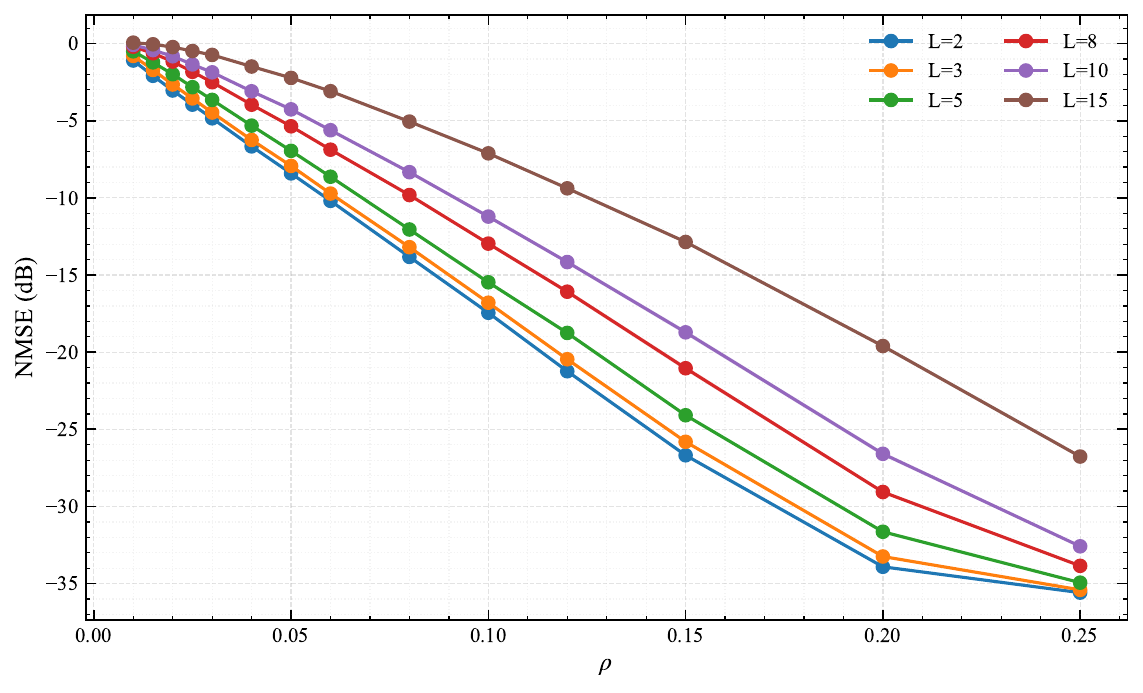}}
	\caption{Recovery curves showing NMSE versus pilot ratio $\rho$
		for varying number of paths $L$. Higher SNR sharpens the
		transition and lowers NMSE within the feasible region.}
	\label{fig:sc_curves}
\end{figure*}

\paragraph*{Recovery curves}
Fig.~\ref{fig:sc_curves} plots NMSE versus pilot ratio $\rho$ for each value of $L$ at $\mathrm{SNR} \in \{10, 20, 30\}$~dB. Two consistent trends emerge. First, the recovery transition shifts toward higher pilot densities as $L$ increases, reflecting the growing intrinsic dimensionality of the channel model. Second, higher SNR sharpens the transition and lowers NMSE once the recovery region is reached. At $\mathrm{SNR} = 10$~dB, no configuration achieves $\mathrm{NMSE} \le 10^{-2}$ within the tested 25\% pilot range, indicating a noise-limited regime where increasing pilot density alone cannot compensate for measurement noise. The curves remain monotonic but exhibit no clear threshold within the tested range.

\begin{figure*}[!t]
	\centering
	\subfloat[SNR = 10 dB]{%
		\includegraphics[width=0.32\textwidth]{%
			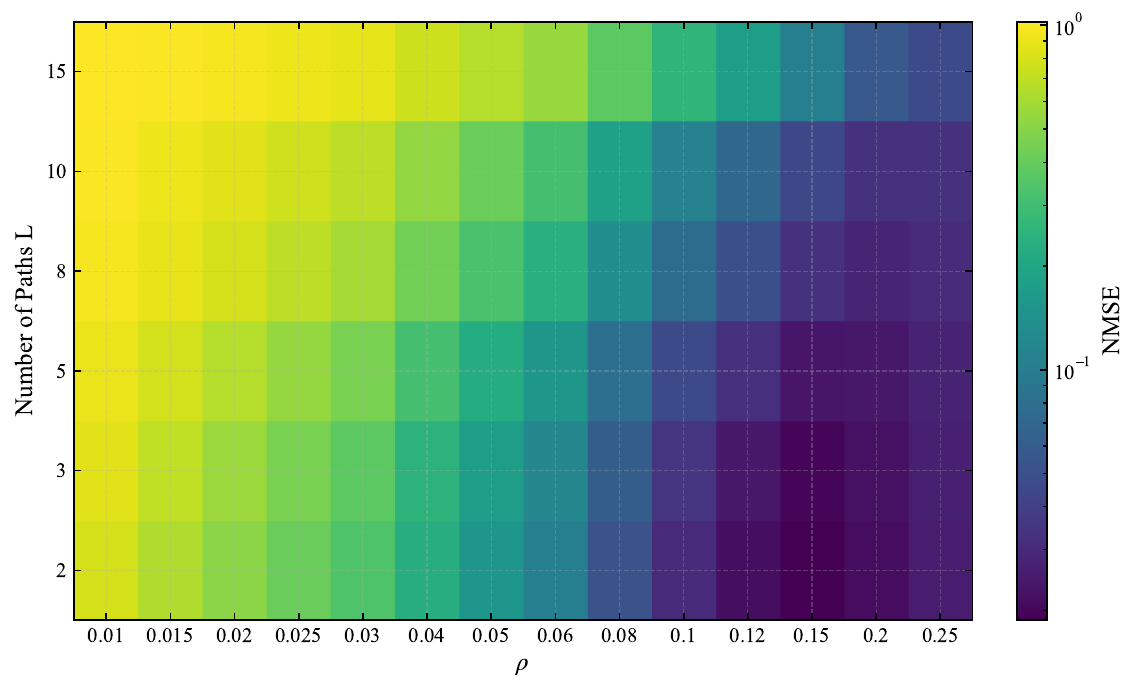}}
	\hfill
	\subfloat[SNR = 20 dB]{%
		\includegraphics[width=0.32\textwidth]{%
			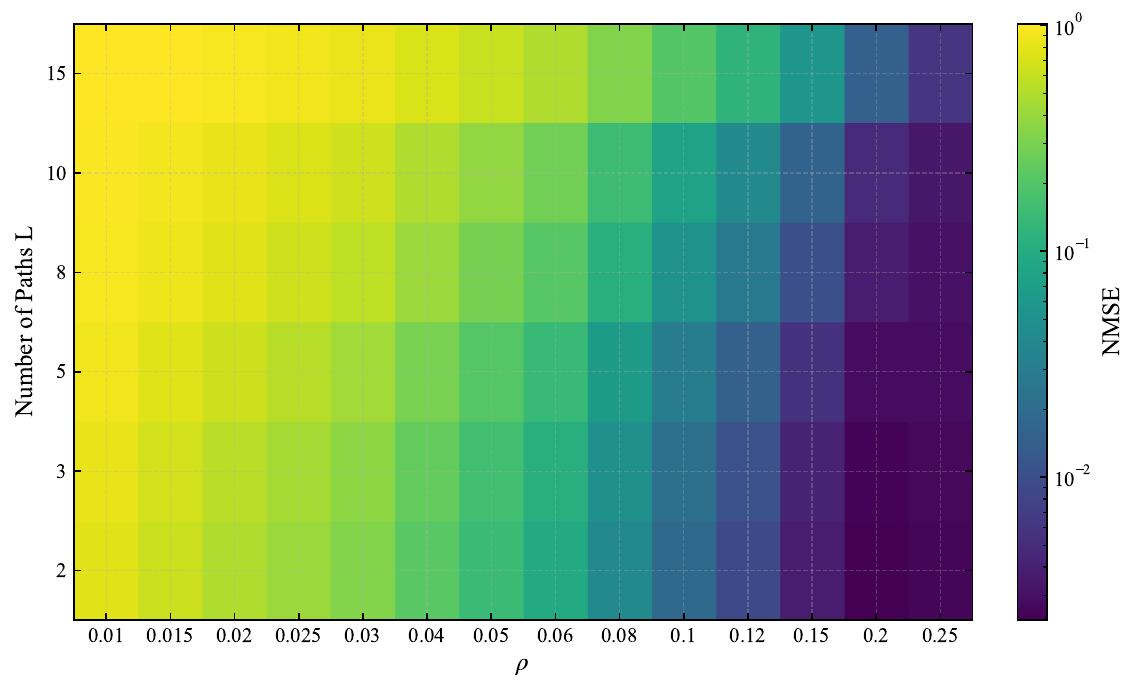}}
	\hfill
	\subfloat[SNR = 30 dB]{%
		\includegraphics[width=0.32\textwidth]{%
			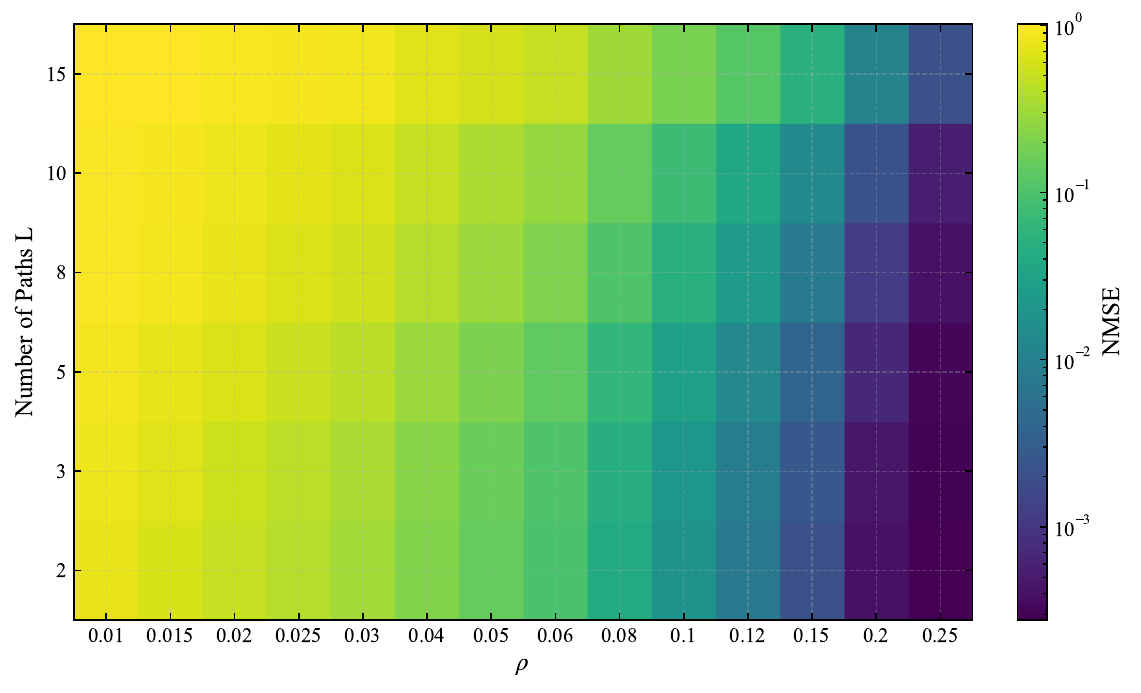}}
	\caption{Recovery region heatmaps over $(L, \rho)$.}
	\label{fig:sc_heatmap}
\end{figure*}

\paragraph*{Recovery region heatmaps}
Fig.~\ref{fig:sc_heatmap} visualizes the recovery region in the $(L, \rho)$ plane. The transition band between successful recovery and failure shifts to higher pilot densities as $L$ increases, consistent with the growth in intrinsic model complexity. At $\mathrm{SNR}=10$~dB, no configuration reaches the $\mathrm{NMSE}\le 10^{-2}$ threshold within the tested range, indicating a noise-limited regime. For $\mathrm{SNR}\in\{20,30\}$~dB, the empirical thresholds lie in the 12--25\% pilot range and increase monotonically with $L$. The difference between 20 and 30~dB is modest (with only small reductions of $\rho_{\min}$ for some $L$), indicating that $L$ dominates the feasibility boundary while higher SNR mainly sharpens the transition and lowers NMSE within the feasible region. These heatmaps provide a compact visualization of the feasibility region for reliable channel reconstruction.

\paragraph*{Empirical thresholds and oversampling factor}
Table~\ref{tab:sc_thresholds} reports the minimum pilot ratio $\rho_{\min}$ and corresponding observation count $|\Omega|_{\min}$ required to achieve $\mathrm{NMSE} \le 10^{-2}$. At $\mathrm{SNR}=10$~dB, no configuration reaches this threshold within the tested pilot range. For $\mathrm{SNR}\in\{20,30\}$~dB, the thresholds fall within the 12--25\% pilot interval and exhibit an overall increasing trend with $L$, indicating that sample complexity is primarily governed by intrinsic channel structure rather than noise level.

\begin{table}[!t]
	\centering
	\caption{Empirical thresholds for $\mathrm{NMSE} \le 10^{-2}$
		and oversampling factor.}
	\label{tab:sc_thresholds}
	\begin{tabular}{@{}ccccc@{}}
		\toprule
		\textbf{SNR (dB)} & $\mathbf{L}$ & $\mathbf{\rho_{\min}}$ &
		$\mathbf{|\Omega|_{\min}}$ & $\mathbf{C}$ \\
		\midrule
		20 &  2 & 0.120 & 15{,}729 & 40.96 \\
		20 &  3 & 0.150 & 19{,}661 & 34.13 \\
		20 &  5 & 0.150 & 19{,}661 & 20.48 \\
		20 &  8 & 0.200 & 26{,}214 & 17.07 \\
		20 & 10 & 0.200 & 26{,}214 & 13.65 \\
		20 & 15 & 0.250 & 32{,}768 & 11.38 \\
		30 &  2 & 0.120 & 15{,}729 & 40.96 \\
		30 &  3 & 0.120 & 15{,}729 & 27.31 \\
		30 &  5 & 0.150 & 19{,}661 & 20.48 \\
		30 &  8 & 0.150 & 19{,}661 & 12.80 \\
		30 & 10 & 0.200 & 26{,}214 & 13.65 \\
		30 & 15 & 0.250 & 32{,}768 & 11.38 \\
		\bottomrule
	\end{tabular}
\end{table}

To interpret these results, this article compares them against the heuristic degrees-of-freedom scaling commonly used in low-rank tensor recovery:
\begin{equation}
	|\Omega|_{\min} \approx C \, L \, (N_r + N_t + N_f),
	\label{eq:dof_scaling}
\end{equation}
where $C$ denotes an oversampling factor capturing deviations from the idealized identifiability limit. This expression is CP-inspired: it counts $L$ factor vectors of lengths $N_r$, $N_t$, and $N_f$, totalling $L(N_r+N_t+N_f)$ free parameters.

For Tucker decompositions with multilinear ranks $(R_r, R_t, R_f)$, the effective degrees of freedom are
\begin{equation}
	d_{\text{Tucker}} = R_r R_t R_f
	+ R_r N_r + R_t N_t + R_f N_f,
	\label{eq:dof_tucker}
\end{equation}
comprising the core tensor ($R_r R_t R_f$ entries) and the three factor matrices ($R_n N_n$ entries each). For the ranks $(R_r, R_t, R_f) = (L, L, L+1)$ used in Experiment~4,
~\eqref{eq:dof_tucker} yields values ranging from $d_{\text{Tucker}} = 524$ at $L=2$ to $d_{\text{Tucker}} = 6608$ at $L=15$, compared with $L(N_r+N_t+N_f) \in
\{384,\,2880\}$ from ~\eqref{eq:dof_scaling}. The ratio $d_{\text{Tucker}}/[L(N_r+N_t+N_f)]$ grows from $\approx\!1.36$ at $L=2$ to $\approx\!2.29$ at $L=15$,
reflecting the cubic growth of the core-tensor term $R_r R_t R_f \sim L^3$ relative to the linear scaling of the CP heuristic.

The rightmost column of Table~\ref{tab:sc_thresholds} lists the empirically observed values of $C$. The measured range ($C \approx 11$--$41$) exceeds theoretical minima established for incoherent low-rank tensors under noiseless and asymptotic conditions~\cite{cai2019nonconvex}. Several practical factors contribute to this gap. First, the CP-inspired reference~\eqref{eq:dof_scaling} increasingly underestimates true Tucker complexity as $L$ grows, since it does not account for the core-tensor degrees of freedom in ~\eqref{eq:dof_tucker}; the empirical $C$ must therefore absorb this structural gap in addition to other sources of inefficiency. Second, finite-sample effects and interpolation in threshold estimation introduce bias. Third, practical convergence limitations of the alternating projection algorithm under noisy and sparse observations may prevent attainment of the optimal recovery boundary.

Nonetheless, the empirical evidence indicates that the minimum observation count scales predominantly with intrinsic channel complexity---characterized by the number of dominant propagation paths and effective subspace dimensions---rather than with the ambient tensor size $N_r N_t N_f$, consistent with Table~\ref{tab:sc_thresholds}.

\paragraph*{Key findings}
The results confirm the existence of a sharp recovery threshold and yield three principal insights:
\begin{enumerate}
	\item Successful tensor completion requires a number of observations that scales with intrinsic model complexity rather than ambient tensor dimension.
	\item The effective oversampling factor $C$ depends on the decomposition type, SNR, and algorithmic efficiency.
	\item At moderate SNR ($\geq 20$~dB), reliable recovery is achievable with 12--25\% pilot density for channels with up to 15 dominant propagation paths.
\end{enumerate}

\subsection{Discussion}
\label{sec:discussion}

The experimental results yield five main findings:

\begin{enumerate}
	
	\item \textbf{Tensor completion mitigates pilot scarcity.}
	Fig.~\ref{fig:exp1_pilots} shows that Tucker completion yields substantial NMSE gains over LS, LMMSE, and OMP across all pilot ratios. At $\rho = 10\%$, Tucker achieves NMSE
	$\approx -11.25$~dB, corresponding to gains of $10.88$~dB over LS ($\approx -0.37$~dB) and $7.83$~dB over OMP ($\approx -3.42$~dB). LS never falls below $-1$~dB across the full sweep, confirming that naive zero-filling cannot recover channel structure under sparse sampling. OMP improves over LS and LMMSE but still exhibits a performance floor around $-3.5$~dB due to dictionary mismatch and pilot scarcity. Notably, at the most extreme sparsity point ($\rho = 2\%$), OMP is slightly better than Tucker by about $0.67$~dB, but Tucker becomes clearly dominant from $\rho \geq 4\%$ onward.
	
	\item \textbf{Decomposition performance depends on channel structure.}
	Fig.~\ref{fig:exp2_cp_tucker} indicates that CP and Tucker perform similarly at low SNR ($-5$~dB: CP $\approx -1.82$~dB versus Tucker $\approx -1.37$~dB, about $0.45$~dB difference), but CP is consistently better across the full SNR range under the synthetic specular model. The CP advantage increases with SNR and reaches $13.11$~dB at SNR\,=\,20~dB (CP $\approx -25.00$~dB versus Tucker $\approx -11.89$~dB), consistent with the specular multipath structure of \eqref{eq:multipath}, where each propagation path contributes a rank-one component that matches the CP parameterization exactly.
	
	Importantly, CP maintains its advantage over Tucker even under the realistic DeepMIMO conditions of Experiment~3: for $\rho \geq 6\%$, CP outperforms Tucker by approximately $1$--$3$~dB across the pilot range. The exception occurs at extreme pilot scarcity ($\rho = 2\%$), where CP exhibits
	heavy-tail divergence (mean NMSE $\approx +13.96$~dB) due to numerical instability of the weighted alternating least-squares algorithm under severe undersampling, while Tucker's alternating-projection scheme remains stable. The hybrid ablation mirrors this same trade-off: Tensor--NN(CP) becomes the most accurate method for $\rho \geq 4\%$, but it still degrades sharply at $\rho = 2\%$ ($\approx +8.03$~dB), whereas Tensor--NN(Tucker) remains stable ($\approx -4.56$~dB). Neural residual learning therefore attenuates, but does not erase, CP's sensitivity at the undersampling boundary; Tucker remains the safer default base when full-range robustness is required.
	
	\item \textbf{Hybrid Tensor--NN improves robustness on realistic channels.}
	Fig.~\ref{fig:exp3_deepmimo} (DeepMIMO, SNR\,=\,10~dB) shows complementary hybrid regimes rather than a single universal winner. Tensor--NN(Tucker) is best at $\rho = 2\%$, improving over raw CP by $18.51$~dB and over ResNet by $1.53$~dB while remaining numerically stable. From $\rho \geq 4\%$, Tensor--NN(CP) becomes the best overall method; at $\rho = 8\%$ it reaches $-16.44$~dB, improving by $4.97$~dB over CP and $2.72$~dB over Tensor--NN(Tucker), and at $\rho = 20\%$ it reaches $-20.34$~dB with a $2.91$~dB margin over CP. This confirms that hybridization inherits the same CP-versus-Tucker trade-off observed in the algebraic baselines: CP-guided residual learning is preferable when the pilot budget is sufficient, whereas Tucker-guided residual learning is preferable at the extreme undersampling boundary.
	
	\item \textbf{Algebraic structure consistently outperforms unconstrained deep learning.}
	The comparison with the ResNet baseline in Fig.~\ref{fig:exp3_deepmimo} shows that Tensor--NN(Tucker) outperforms ResNet at every pilot density tested, with a monotonically growing margin: $1.53$~dB at $\rho = 2\%$, $3.64$~dB at $\rho = 4\%$, $3.58$~dB at $\rho = 6\%$, $3.67$~dB at $\rho = 8\%$, $3.91$~dB at $\rho = 10\%$, $4.91$~dB at $\rho = 15\%$, and $5.67$~dB at $\rho = 20\%$. Tensor--NN(CP) widens this gap once the CP prior becomes stable, but its behavior at $\rho = 2\%$ confirms that structural accuracy alone is insufficient without numerical robustness. The global SNR reference of \eqref{eq:snr_def} ensures that these advantages are not artefacts of the measurement setting. Taken together, the results demonstrate that the algebraic prior is the dominant source of performance gain, and that the neural residual stage provides complementary correction for model mismatch rather than compensating for insufficient structure.
	
	\item \textbf{Sample complexity is governed by intrinsic channel structure.}
	Experiment~4 shows that the minimum pilot ratio $\rho_{\min}$ and observation count $|\Omega|_{\min}$ increase monotonically with the number of propagation paths $L$. At SNR\,=\,10~dB, no configuration reaches $\mathrm{NMSE} \leq 10^{-2}$ within the tested range, indicating a noise-limited regime where increasing pilot density alone cannot compensate for measurement noise. For SNR\,$\in$\,$\{20,\,30\}$~dB, recovery thresholds lie in the 12--25\% pilot range and are primarily governed by $L$ rather than by the ambient tensor dimension $N_r N_t N_f$, consistent with Table~\ref{tab:sc_thresholds}.
	
\end{enumerate}

\subsection{Limitations}
\label{subsec:limitations}

Despite the gains reported above, the proposed framework has limitations that merit explicit discussion:

\begin{itemize}
	\item \textbf{Rank selection and sensitivity.} The tensor methods require multilinear ranks (or CP rank) specified a priori. The current implementation fixes ranks per experiment and, in the sample-complexity study, ties them to the number of paths via $(R_r, R_t, R_f) = (L, L, L{+}1)$. Performance degrades under rank mismatch, and a principled, data-driven rank selection strategy remains necessary for robust deployment.
	
	\item \textbf{High-mobility and time-varying channels.} The experiments assume quasi-static channels and single-snapshot estimators. Rapid mobility, Doppler spread, and temporal correlation fall outside the model; extending the framework to spatio-temporal estimation or channel tracking remains future work.
	
	\item \textbf{Neural training requirements.} The neural components (Tensor--NN residual and ResNet baselines) train offline on paired $(\mathcal{H}, \mathcal{Y}_\Omega)$ data under multi-SNR and multi-pilot regimes. Deployment may require representative training data (or synthetic surrogates) and fine-tuning to mitigate distribution shift across scenarios, environments, or array configurations. Although the residual network is fully convolutional and can be re-applied to different tensor sizes without architectural change, cross-scenario transfer and strict spatial generalization without retraining are not claimed here.
	
	\item \textbf{Synchronization and hardware impairments.} The observation model covers only entry-wise sampling with additive Gaussian noise. Effects such as carrier-frequency offset, phase noise, in-phase/quadrature (IQ) imbalance, nonlinear power amplifiers, low-resolution quantization, and other hardware imperfections fall outside the model. Their impact on tensor completion and neural residual learning remains an open question.
\end{itemize}

\subsection{Reproducibility}
\label{sec:reproducibility}

The reference implementation was executed under Python~3.10 with PyTorch~2.1. The wall-clock timings reported in Table~\ref{tab:runtime_complexity} were measured on a MacBook Pro with Apple M3 Pro (11 CPU cores, 14 graphics processing unit (GPU) cores, 18~GB random-access memory (RAM)) using the MPS backend. Because runtime depends on hardware and backend, these values should be interpreted comparatively rather than as absolute deployment guarantees. A fixed random seed (\texttt{base\_seed=42}) ensures deterministic channel generation, pilot sampling, and network initialization across all experiments.

The complete codebase---including data generation scripts, training configurations, pretrained model checkpoints, and plotting utilities---is publicly available at:
\begin{center}
	\url{https://github.com/alexandreblima/MIMO-TN}
\end{center}
The experiments generate DeepMIMO channels using the official dataset API~\cite{alkhateeb2019deepmimo} with configuration files provided in the repository. Researchers can reproduce all figures and tables by executing the provided shell scripts with default parameters.

\section{Conclusion}
\label{sec:conclusion}

This paper proposed a structure-informed framework for pilot-limited wideband MIMO channel estimation combining low-rank tensor completion with neural residual learning. The formulation directly addresses the ill-posed inverse problem induced by sparse pilot observations by exploiting algebraic priors that arise universally from sparse multipath propagation and array geometry, without requiring site-specific environmental models. SNR is defined globally with respect to the average power per channel coefficient over the full tensor---independent of the pilot set $\Omega$ and pilot ratio $\rho$---ensuring strict comparability across all experiments and pilot densities.

Synthetic-channel experiments demonstrate that Tucker completion achieves $10.88$~dB NMSE improvement over LS and $7.83$~dB over OMP at $\rho = 10\%$ pilot density.
CP outperforms Tucker across the full SNR range under the specular multipath model, reaching a $13.11$~dB advantage at SNR\,=\,20~dB, consistent with the rank-one multipath structure that matches the CP parameterization exactly. On DeepMIMO ray-tracing channels, CP maintains its advantage over Tucker for $\rho \geq 6\%$ but exhibits severe heavy-tail divergence at $\rho = 2\%$ (mean NMSE $\approx +13.96$~dB), where Tucker's alternating-projection scheme remains numerically stable. This same trade-off persists in the hybrid estimators: Tensor--NN(Tucker) remains robust at $\rho = 2\%$ ($\approx -4.56$~dB), whereas Tensor--NN(CP) becomes the best-performing estimator from $\rho \geq 4\%$, reaching $-16.44$~dB at $\rho = 8\%$ and $-20.34$~dB at $\rho = 20\%$, with gains of $2.72$~dB and $1.90$~dB over Tensor--NN(Tucker), respectively. Tensor--NN(Tucker) still outperforms the purely data-driven ResNet baseline at every pilot density tested, confirming that algebraic structure remains the dominant source of performance gain, while CP-guided residual learning provides an additional accuracy boost once the CP prior is numerically stable.

Empirical recovery thresholds confirm that the minimum observation count scales primarily with intrinsic channel complexity---governed by the number of dominant propagation
paths $L$---rather than with the ambient tensor dimension $N_r N_t N_f$. At SNR\,$\geq$\,20~dB, channels with up to 15 dominant paths admit reliable recovery with $12$--$25\%$ pilot density.

These findings establish tensor priors as physically grounded and theoretically motivated constraints for pilot-limited estimation, while neural residual learning
provides a principled mechanism to compensate model mismatch under realistic propagation conditions. The fully convolutional residual network can be instantiated for different tensor sizes at the architectural level, but cross-scenario transfer and strict spatial generalization without retraining remain open problems. Future work will address adaptive rank selection via model-order estimation, extensions to time-varying channels through joint space--frequency--time tensor modeling, integration with hybrid analog-digital architectures, and robustness analysis under hardware impairments including phase noise, nonlinearities, and low-resolution quantization.

\end{document}